\documentclass[fleqn,usenatbib]{mnras}

\usepackage{newtxtext,newtxmath}

\usepackage[T1]{fontenc}

\usepackage{graphicx}	
\usepackage{amsmath}	
\usepackage{booktabs}
\usepackage{xcolor}
\usepackage{multirow}

\newcommand{\galfit}{\textsc{Galfit}}
\newcommand{\Msun}{M_\odot}
\newcommand{\reff}{R_{\mathrm{eff}}}
\newcommand{\reffi}{R_{\mathrm{eff}, i}}
\newcommand{\Mobsi}{M_{\mathrm{obs}, i}}
\newcommand{\rmax}{R_{\mathrm{max}}}
\newcommand{\Msunpc}{\,\Msun\,\mathrm{pc}}
\newcommand{\SigmaGMC}{\Sigma_{\mathrm{GMC}}}

\defcitealias{ryon_etal_17}{R17}
\defcitealias{hogg_etal_10}{H10}
\defcitealias{gieles_renaud_16}{GR16}

\graphicspath{{./}{figures/}}

\title[Star Cluster Radii]{Radii of Young Star Clusters in Nearby Galaxies}

\author[G. Brown, O. Y. Gnedin]{
Gillen Brown\href{https://orcid.org/0000-0002-9114-5197}{\includegraphics[scale=0.04]{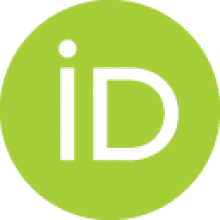}}\thanks{E-mail: gillenb@umich.edu} and
Oleg Y. Gnedin\href{https://orcid.org/0000-0001-9852-9954}{\includegraphics[scale=0.04]{orcid.png}}
\\
Department of Astronomy, University of Michigan, Ann Arbor, MI 48109, USA\\
}

\date{Accepted XXX. Received YYY; in original form ZZZ}

\pubyear{2021}

\begin{document}
\label{firstpage}
\pagerange{\pageref{firstpage}--\pageref{lastpage}}
\maketitle

\begin{abstract}
We measure the projected half-light radii of young star clusters in 31 galaxies from the Legacy Extragalactic UV Survey (LEGUS). We implement a custom pipeline specifically designed to be robust against contamination, which allows us to measure radii for 6097 clusters. This is the largest sample of young star cluster radii currently available. We find that most (but not all) galaxies share a common cluster radius distribution, with the peak at around 3~pc. We find a clear mass-radius relation of the form $\reff \propto M^{0.24}$. This relation is present at all cluster ages younger than 1~Gyr, but with a shallower slope for clusters younger than 10~Myr. We present simple toy models to interpret these age trends, finding that high-mass clusters are more likely to be not tidally limited and expand. We also find that most clusters in LEGUS are gravitationally bound, especially at older ages or higher masses. Lastly, we present the cluster density and surface density distributions, finding a large scatter that appears to decrease with cluster age. The youngest clusters have a typical surface density of 100$\Msunpc^{-2}$.
\end{abstract}

\begin{keywords}
galaxies: star clusters: general -- galaxies: star formation
\end{keywords}

\section{Introduction} \label{sec:intro}

Young star clusters bridge the small scales of star formation and the large scales of galaxy formation. They are easily detected in nearby star-forming galaxies and contain the majority of massive stars \citep{krumholz_etal_19_review}. Their almost universal luminosity function, and corresponding mass function \citep{adamo_etal20}, is used both as a test of molecular cloud collapse models \citep{ballesteros_etal20} and as a constraint on the star formation modeling in cosmological simulations \citep{li_etal18}. However, our understanding of the origin and long-term evolution of star clusters is still hindered by lack of reliable measurements of their density distribution. 

Depending on the initial density, young massive clusters may evolve into globular clusters or may dissolve into the smooth stellar field \citep{portegies_zwart_etal_10}. As clusters form in giant molecular clouds (GMCs), feedback from massive stars ejects the residual gas, making the cluster expand to re-establish virial equilibrium \citep{goodwin_bastian_06}. Additionally, tidal shocks from encounters with other GMCs or spiral arms kinematically heat the cluster and lead to its disruption \citep{spitzer_1958}. The density of young clusters determines whether they can survive this intense phase as bound stellar systems. 

To calculate the density of young star clusters, one needs to measure their radii. Currently published measurements are limited to only a few galaxy samples. \citet{kharchenko_etal_13} measured the radii for $\sim1100$ open clusters in the Milky Way, while samples from external galaxies include \citet{bastian_etal_12} and \citet{ryon_etal_15} who measured radii for several hundred clusters in M83, about 3800 clusters in M51 measured by \citet{chandar_etal16}, and 514 clusters in M31 measured by the PHAT survey \citep{johnson_etal_12,fouesneau_etal_14}. Of particular note for this paper, \citet{ryon_etal_17} (hereafter R17) measured the radii of several hundred clusters spread between two galaxies in the Legacy Extragalactic UV Survey (LEGUS). 

In this paper we measure the projected half light radius (effective radius) of clusters in the 31 galaxies with publicly available cluster catalogs from LEGUS. Our method for fitting the radii is described in Section~\ref{sec:methods}. In Section~\ref{sec:results} we describe our findings of a cluster radius distribution common to most galaxies and a clear cluster mass-radius relation. In Section~\ref{sec:discussion} we discuss how selection effects affect our results, calculate cluster densities, and present a toy model of cluster evolution. We close with a summary in Section~\ref{sec:conclusions}.

\section{Methods} \label{sec:methods}

\subsection{The LEGUS sample}

We use the publicly available LEGUS dataset\footnote{\url{https://archive.stsci.edu/prepds/legus/dataproducts-public.html}} to extend the sample of clusters with uniformly measured radii and densities to the 31 galaxies with currently available cluster catalogs. We summarize some of the key details of LEGUS in this section (see \citealt{calzetti_etal_15_legus} for more on the LEGUS survey description, \citealt{Adamo_etal_17} and \citealt{cook_etal_19} for more on the cluster catalogs).

LEGUS is a Cycle 21 Treasury program on HST that collected imaging with WFC3/UVIS to supplement archival ACS/WFC imaging, producing five-band coverage from the near-UV to the I band for 50 galaxies. Within each field, a uniform process was used to identify cluster candidates. 

First, SourceExtractor \citep{bertin_arnouts_96} is used to find sources in the white-light images (a combination of the images in all five filters, weighted by S/N). Next, a user selects a training set of objects that are clearly clusters or stars. The pipeline calculates  the concentration index (CI), which is the magnitude difference between apertures of radius 1 pixel and 3 pixels \citep{holtzman_etal_92,whitmore_etal_10}.  The user then selects a value of the CI that separates stars from clusters. The aperture for science photometry is chosen as the integer number of pixels containing at least 50\% of the cluster flux, based on the curve of growth of the clusters. This science photometry is done with the same aperture in all filters, using a sky annulus located at 7 pixels with a width of 1 pixel. 

Next, aperture corrections for each filter are determined using an averaged method and CI-based method. In this study we use catalogs using the averaged aperture correction method, so we describe that here. The correction is the difference between the magnitude of the source obtained using an aperture of 20 pixels and the magnitude obtained from the science aperture. The average aperture correction for a user-defined set of well-behaved clusters is used for all clusters. This photometry is corrected for galactic foreground extinction \citep{schlafly_finkbeiner_11}. Sources that are detected in at least four filters with a photometric error of less than 0.3 mag, have an absolute V-band magnitude brighter than $-6$, and have a larger CI than the limit determined earlier, are then visually inspected by the LEGUS team. Sources that do not pass some of these cuts can be manually added by LEGUS team members, but the number is small. 

Three or more team members visually inspect each cluster candidate, classifying it into one of the following four classes. Class 1 objects are compact and centrally concentrated with a homogeneous color. Class 2 clusters have slightly elongated density profiles and a less symmetric light distribution. Class 3 clusters are likely compact associations, having asymmetric profiles or multiple peaks on top of diffuse underlying wings. Class 4 objects are stars or artifacts. 

The age, mass, and extinction of each cluster are determined by the SED fitting code \textsl{Yggdrasil} \citep{zackrisson_etal_11}. Versions of the catalogs are created with different stellar tracks and extinction laws, but in this paper we select the catalogs that use the MW extinction law \citep{cardelli_etal_89} and Padova-AGB tracks available in Starburst99 \citep{leitherer_etal_99_starburst,vasquez_leitherer_05_starburst}.

The survey targeted 50 galaxies. Currently, public cluster catalogs are available for 31 of these galaxies. Some galaxies are observed with multiple fields, resulting in 34 total fields with cluster catalogs. Table~\ref{tab:galaxies} lists these fields along with some key properties of the galaxies. 

\subsection{Outline of the measurement procedure}

To fit the cluster radii, we implement a custom pipeline written in Python. We choose to implement our own method to have full control over the fitting process and better quantify the distribution of errors of the fit parameters. It is analogous to that in the popular package \galfit\ \citep{peng_etal02_galfit,peng_etal10_galfit}, but with automated masking and several other features described in Section~\ref{sec:fitting} that make it more robust against contamination from other nearby sources and ensure a good estimate of the local background. We assume an EFF density profile for young clusters \citep{elson_fall_freeman_87,larsen_99,McLaughlin_vanderMarel_05}, then convolve it with the empirically-measured point spread function (PSF) before comparing to the data.

\begin{figure}
    \vspace{2mm}
    \includegraphics[width=\columnwidth]{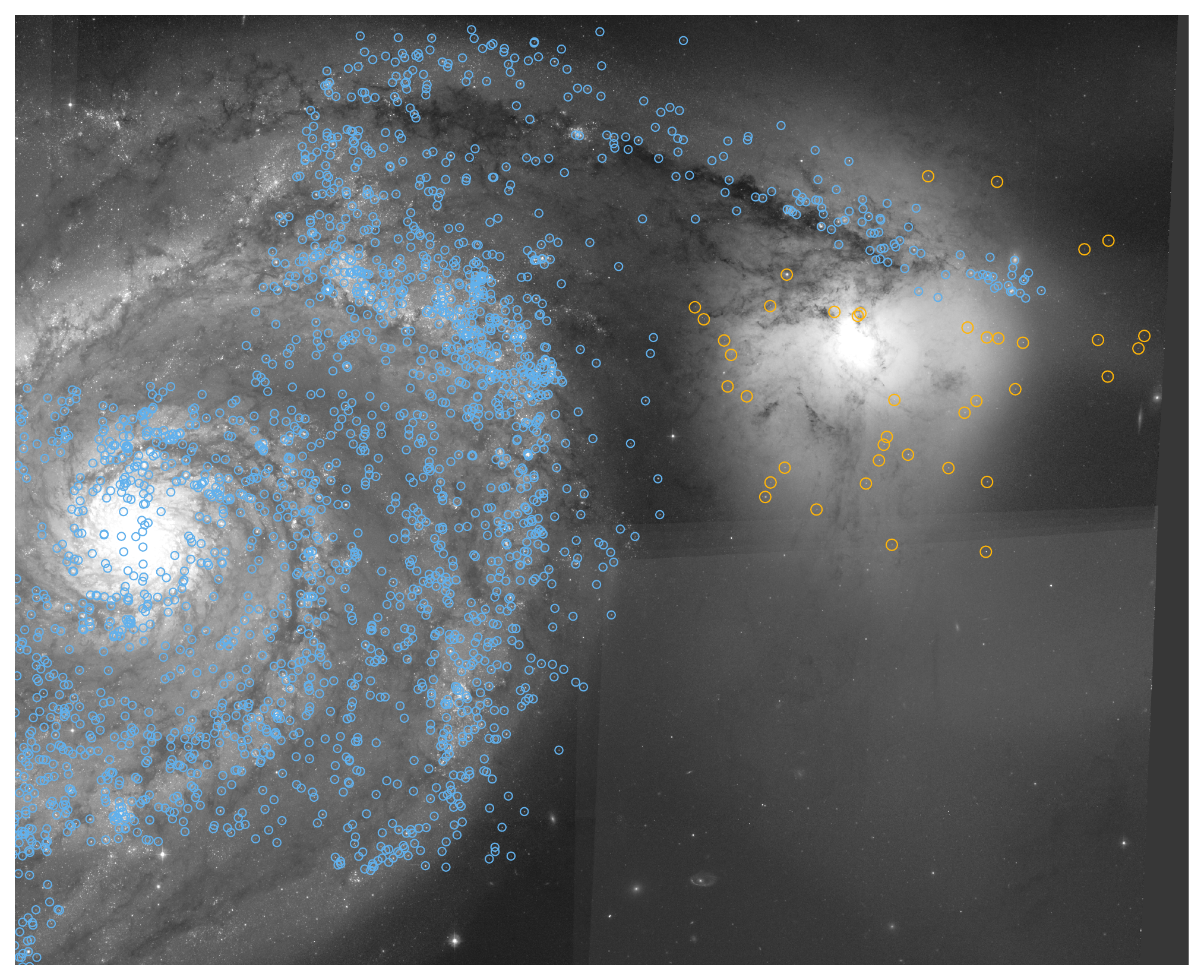}
    \vspace{-6mm}
    \caption{A portion of the F555W image of the LEGUS field with NGC~5194 and NGC~5195, illustrating our selection of cluster belonging to NGC~5194 (blue circles) and NGC~5195 (orange circles).}
    \label{fig:ngc5194_5195}
\end{figure}

We use the F555W images in 25 of the 34 fields, but for the other 9 fields that were not observed in F555W, we instead use F606W. All images have a pixel scale of 39.62 mas pixel$^{-1}$. The LEGUS mosaics have units of e$^{-}$s$^{-1}$, and we multiply by the exposure time to convert to the electron count. We use the recommended LEGUS cluster catalogs that adopt the MW extinction, Padova stellar evolutionary tracks, and the averaged aperture correction method \citep{Adamo_etal_17}. The LEGUS team created visual classification tags by visually inspecting each cluster with multiple team members. We select clusters that were identified as being concentrated and either symmetric (class 1) or with some degree of asymmetry (class 2). We use the mode of the classifications from multiple team members. Additionally, machine learning classifications are available for several galaxies \citep{grasha_18, grasha_etal_19}. For NGC~5194 and NGC~5195, we use the human classifications for clusters where those are available, and supplement with machine learning classifications for clusters not inspected by humans. In NGC~1566, we use the hybrid classification system created by the LEGUS team, where some clusters are inspected by humans only, some by machine learning only, and some with a machine learning classification verified by humans. We do not use the machine learning classifications for NGC~4449, as we find that the machine-classified clusters have significantly different concentration index and radius distributions from the rest of the sample. This selection produces a final sample of 7242 clusters. As NGC~5194 and NGC~5195 are in the same LEGUS field, we manually identify their host galaxy. Figure~\ref{fig:ngc5194_5195} shows this selection.

\begin{table*}
    \centering
    \caption{List of the LEGUS fields included in the sample, with some key properties of their cluster populations. We use the TRGB based distances from \citet{sabbi_etal18} except for NGC~1566 -- see the discussion at the end of Section \ref{sec:fitting}. Stellar masses and specific star formation rates are from \citet{calzetti_etal_15_legus}, who obtained SFR from \textit{GALEX} far-UV corrected for dust attenuation, as described in \citet{lee_etal_09}, and stellar mass from extinction-corrected B-band luminosity and color information, as described in \citet{bothwell_etal_09} and using the mass-to-light ratio models of \citet{bell_dejong_01}. Note that NGC~5194 and NGC~5195 are in the same field, so they share a distance and PSF, but we split them for clarity.}
	\begin{tabular}{lrccccc}
		\toprule
		LEGUS Field & N & log $M_\star$ ($\Msun$) & log sSFR (yr$^{-1}$) & Distance (Mpc) & PSF size (pc) & Cluster $\reff$: 25--50--75th percentiles \\ 
		\midrule
		IC 4247 & 5 & 8.08 & -10.18 & 5.11 $\pm$ 0.4 & 1.48 & 2.30 -- 3.52 -- 4.36 \\ 
		IC 559 & 21 & 8.15 & -10.45 & 10.0 $\pm$ 0.9 & 2.74 & 4.05 -- 4.85 -- 6.34 \\ 
		NGC 1313-e & 137 & 9.41 & -9.35 & 4.30 $\pm$ 0.24 & 1.26 & 1.41 -- 2.35 -- 3.00 \\ 
		NGC 1313-w & 276 & 9.41 & -9.35 & 4.30 $\pm$ 0.24 & 1.26 & 1.13 -- 2.54 -- 3.58 \\ 
		NGC 1433 & 112 & 10.23 & -10.80 & 9.1 $\pm$ 1.0 & 2.53 & 1.11 -- 1.79 -- 3.20 \\ 
		NGC 1566 & 881 & 10.43 & -9.68 & 15.6 $\pm$ 0.6 & 4.33 & 3.00 -- 4.30 -- 6.28 \\ 
		NGC 1705 & 29 & 8.11 & -9.07 & 5.22 $\pm$ 0.38 & 1.35 & 2.73 -- 3.26 -- 4.20 \\ 
		NGC 3344 & 237 & 9.70 & -9.76 & 8.3 $\pm$ 0.7 & 2.37 & 1.72 -- 2.40 -- 3.53 \\ 
		NGC 3351 & 19 & 10.32 & -10.13 & 9.3 $\pm$ 0.9 & 2.75 & 0.88 -- 2.55 -- 5.17 \\ 
		NGC 3738 & 142 & 8.38 & -9.54 & 5.09 $\pm$ 0.40 & 1.52 & 2.17 -- 3.25 -- 4.30 \\ 
		NGC 4242 & 12 & 9.04 & -10.04 & 5.3 $\pm$ 0.3 & 1.54 & 1.28 -- 2.47 -- 3.11 \\ 
		NGC 4395-n & 20 & 8.78 & -9.25 & 4.54 $\pm$ 0.18 & 1.24 & 1.34 -- 1.84 -- 2.42 \\ 
		NGC 4395-s & 95 & 8.78 & -9.25 & 4.54 $\pm$ 0.18 & 1.33 & 0.49 -- 0.82 -- 1.76 \\ 
		NGC 4449 & 425 & 9.04 & -9.07 & 4.01 $\pm$ 0.30 & 1.17 & 1.68 -- 2.40 -- 3.32 \\ 
		NGC 45 & 14 & 9.52 & -9.97 & 6.8 $\pm$ 0.5 & 1.84 & 2.41 -- 3.25 -- 4.61 \\ 
		NGC 4656 & 184 & 8.60 & -8.90 & 7.9 $\pm$ 0.7 & 2.11 & 2.35 -- 3.32 -- 4.15 \\ 
		NGC 5194 & 2921 & 10.38 & -9.54 & 7.40 $\pm$ 0.42 & 2.16 & 1.29 -- 2.17 -- 3.29 \\ 
		NGC 5195 & 40 & 10.36 & -10.82 & 7.40 $\pm$ 0.42 & 2.16 & 2.69 -- 3.39 -- 4.88 \\ 
		NGC 5238 & 8 & 8.15 & -10.15 & 4.43 $\pm$ 0.34 & 1.29 & 2.29 -- 2.63 -- 2.94 \\ 
		NGC 5253 & 57 & 8.34 & -9.34 & 3.32 $\pm$ 0.25 & 0.98 & 1.09 -- 1.86 -- 2.66 \\ 
		NGC 5474 & 143 & 8.91 & -9.48 & 6.6 $\pm$ 0.5 & 1.95 & 2.28 -- 3.25 -- 4.13 \\ 
		NGC 5477 & 14 & 7.60 & -9.12 & 6.7 $\pm$ 0.5 & 1.98 & 2.40 -- 3.64 -- 4.78 \\ 
		NGC 628-c & 691 & 10.04 & -9.48 & 8.8 $\pm$ 0.7 & 2.59 & 1.70 -- 2.43 -- 3.44 \\ 
		NGC 628-e & 172 & 10.04 & -9.48 & 8.8 $\pm$ 0.7 & 2.49 & 1.95 -- 2.64 -- 4.05 \\ 
		NGC 6503 & 167 & 9.28 & -9.77 & 6.3 $\pm$ 0.5 & 1.69 & 1.31 -- 2.12 -- 3.34 \\ 
		NGC 7793-e & 108 & 9.51 & -9.79 & 3.79 $\pm$ 0.20 & 0.92 & 0.60 -- 1.02 -- 2.20 \\ 
		NGC 7793-w & 135 & 9.51 & -9.79 & 3.79 $\pm$ 0.20 & 1.11 & 0.66 -- 1.26 -- 2.20 \\ 
		UGC 1249 & 48 & 8.74 & -9.56 & 6.4 $\pm$ 0.5 & 1.88 & 1.83 -- 2.58 -- 3.26 \\ 
		UGC 4305 & 45 & 8.36 & -9.28 & 3.32 $\pm$ 0.25 & 0.97 & 0.69 -- 1.15 -- 1.91 \\ 
		UGC 4459 & 7 & 6.83 & -8.99 & 3.96 $\pm$ 0.30 & 1.23 & 1.83 -- 2.51 -- 3.13 \\ 
		UGC 5139 & 9 & 7.40 & -9.10 & 3.83 $\pm$ 0.29 & 1.12 & 1.07 -- 1.54 -- 4.00 \\ 
		UGC 685 & 11 & 7.98 & -10.13 & 4.37 $\pm$ 0.34 & 1.31 & 1.77 -- 2.23 -- 2.96 \\ 
		UGC 695 & 11 & 8.26 & -9.95 & 7.8 $\pm$ 0.6 & 2.08 & 1.39 -- 6.21 -- 7.49 \\ 
		UGC 7408 & 35 & 7.67 & -9.67 & 7.0 $\pm$ 0.5 & 2.08 & 3.75 -- 4.41 -- 5.68 \\ 
		UGCA 281 & 11 & 7.28 & -8.98 & 5.19 $\pm$ 0.39 & 1.51 & 3.12 -- 3.50 -- 4.25 \\ 
		\midrule
		Total & 7242 & -- & -- & -- & --  & 1.53 -- 2.48 -- 3.69 \\ 
		\bottomrule
	\end{tabular}
    \label{tab:galaxies}
\end{table*}

\subsection{PSF creation} \label{sec:psf}

\begin{figure}
    \vspace{2mm}
    \includegraphics[width=\columnwidth]{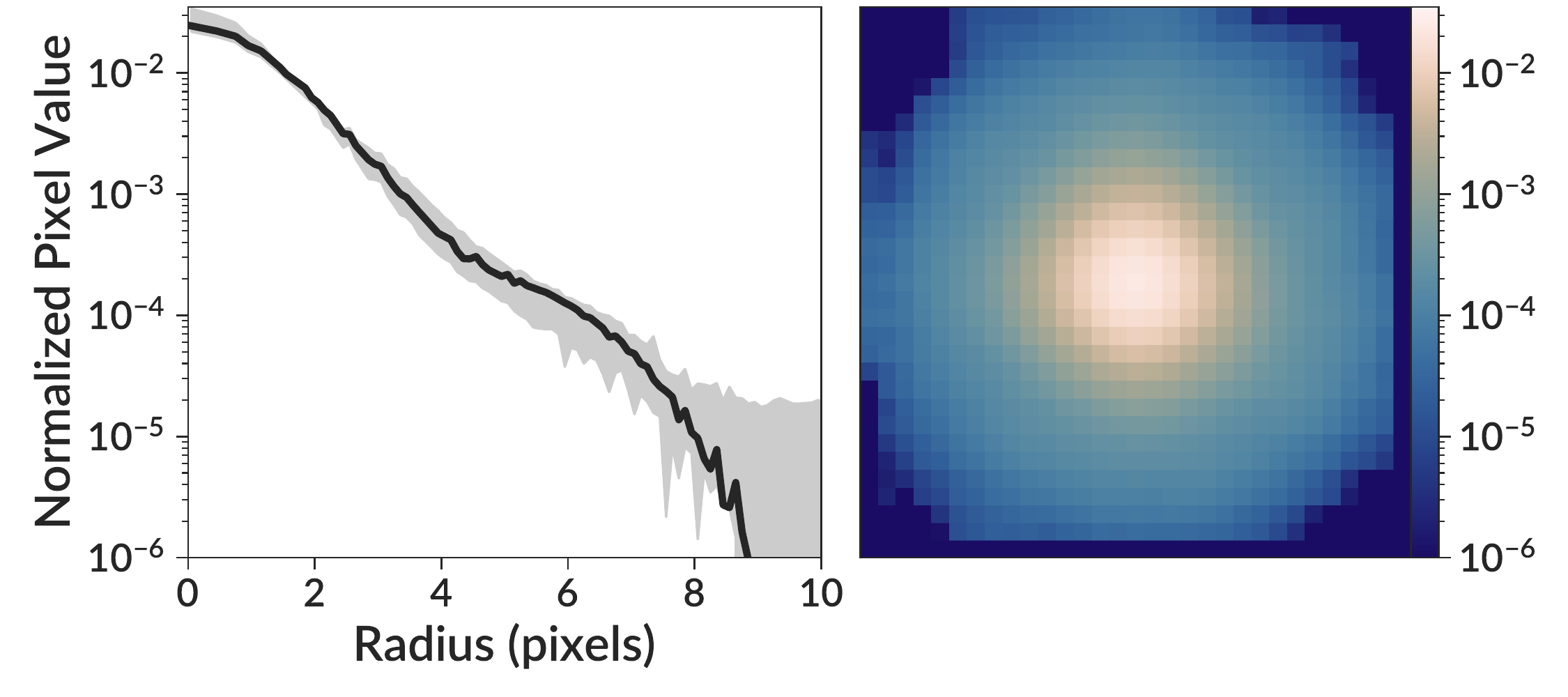}
    \vspace{-6mm}
    \caption{Visualization of one of our PSFs, from the NGC~1313-e field. The solid line in the left panel shows the azimuthally-averaged radial profile of this PSF, while the shaded region shows the range of the PSF profiles from all other fields, normalized to integrate to unity. The right panel shows an image of the NGC~1313-e PSF.}
    \label{fig:psf}
\end{figure}

Like \galfit, our method convolves the PSF with the model image, and compares the result with the observed data. To produce this PSF, we use \texttt{Photutils} \citep{photutils}, an Astropy package \citep{astropy_i,astropy_ii}. We manually select bright isolated stars in each field, then use the \texttt{EPSFBuilder} class of \texttt{Photutils} to create a separate PSF for each field. \texttt{EPSFBuilder} follows the prescription of \citet{anderson_etal_00}. The final PSF images are 15 pixels (0.59$\arcsec$) wide, and we spatially subsample the PSF by a factor of two, producing a PSF with twice the spatial resolution of the input image. We do not choose higher values, as this significantly increases the computational cost of the fitting procedure (particularly the convolution). As shown in Figure~\ref{fig:ryon_comparison}, our results are consistent with those of \citetalias{ryon_etal_17} for NGC~1313 and NGC~628, even for the smallest clusters, indicating that this oversampling factor is adequate. We use \texttt{Photutils}'s ``quadratic'' smoothing kernel, which is the polynomial fit with degree=2 to 5x5 array of zeros with 1 at the center. We found that the other options gave unphysically non-smooth PSFs. Figure~\ref{fig:psf} illustrates our created PSFs.

\subsection{Fitting cluster parameters} \label{sec:fitting}

We fit the clusters with the EFF surface brightness profile, as it accurately describes the light profiles of young star clusters \citep{elson_fall_freeman_87,larsen_99,McLaughlin_vanderMarel_05,bastian_etal_12,ryon_etal_15, ryon_etal_17, cuevasotahola_etal_20}. Assuming circular symmetry, it takes the form
\begin{equation}
    \mu(r) = \mu_0 \left( 1 + \frac{r^2}{a^2}\right)^{-\eta}
\end{equation}
where $\mu$ is the surface brightness, $a$ is the scale radius, and $\eta$ is the power law slope. As real clusters are typically not circularly symmetric, we include ellipticity as follows:
\begin{equation}
    \mu(x, y) = \mu_0 \left( 1 + \left[\frac{x'(x, y)}{a}\right]^2 + \left[\frac{y'(x, y)}{aq}\right]^2 \right)^{-\eta}
    \label{eq:mu}
\end{equation}
where $q$ is the ratio of the minor to major axes of the ellipse ($0 < q \leq 1$). We have rotated the image coordinate system by angle $\theta$ about the cluster center ($x_c, y_c$) to new coordinates ($x'$, $y'$) as follows:
\begin{align}
    x'(x, y) &=  (x-x_c) \cos \theta + (y-y_c) \sin \theta \\
    y'(x, y) &= -(x-x_c) \sin \theta + (y-y_c) \cos \theta.
\end{align}
Here $x'$ is aligned with the cluster major axis, while $y'$ is aligned with the minor axis. This gives 7 cluster parameters: $\mu_0$, $x_c$, $y_c$, $a$, $q$, $\theta$, $\eta$. We also leave the local background  $f_{\rm BG}$ as a free parameter, giving 8 total parameters to fit.

We perform this fit on a 30$\times$30 pixel snapshot centered on the cluster, following \citetalias{ryon_etal_17}. We tested larger snapshot sizes (40 and 50 pixels) but found no significant differences in fitted cluster radii, even for the biggest clusters where a larger snapshot could potentially allow for a better determination of the local background. These larger snapshot sizes also included more contaminating sources, leading to more catastrophic fit failures. The 30-pixel snapshot minimizes these failures while still performing well on the largest clusters.

To account for contaminating sources inside this 30$\times$30 pixel snapshot, we mask star-like sources identified by the \texttt{IRAFStarFinder} class of \texttt{Photutils}. Any pixels within 2$\times$FWHM of the stars are masked. However we discard any stars whose masked region would extend within 3 pixels of the cluster center, as well as any sources with a peak pixel value less than 2 times the local sky background identified by \texttt{IRAFStarFinder}. This second criterion was added to stop the masking of substructure in the most extended clusters. We also mask any pixels that are within 6 pixels of another star cluster, in cases where two star clusters are close to each other.

However, an automated masking system cannot solve all issues with contamination. To make our fitting method robust to potential contamination, we have made substantial modifications compared to a \galfit-like method.

Our best-fit parameters maximize the posterior distribution, defined as
\begin{equation}
    \log{P_{\rm posterior}} = -\frac{1}{2}\sum_{x,y} w(x, y) \left|\frac{f_d(x, y) - f_m(x, y)}{\sigma(x, y)}\right| + \log{P_{\rm prior}}
    \label{eq:likelihood}
\end{equation}
where $x$ and $y$ are pixel coordinates, $w$ are pixel weights, $f_d$ is the data value at pixel $(x,y)$, $f_m$ is the model, $\sigma(x, y)$ is the pixel uncertainty, and $P_{\rm prior}$ is the prior distribution. We will expand on each of these components in turn. Note that we use the absolute value of the differences between the model and data rather than the more typical square. As the square weights large differences more heavily, it has the effect of increasing the attention the fit pays to unmasked contamination, as these pixels have large deviations. Using the absolute value instead produces fits that are less affected by contamination.

In addition, pixel weights are used to reduce the effect of contamination, particularly at large distances from the cluster. We weight each pixel proportional to $1/r$ so that each annulus has the same weight. To avoid giving dramatically more weight to the most central pixels, all pixels within 3 pixels from the center receive the same weight. We use the distance from the center of the cluster to determine the radius, giving $r^2 = (x-x_c)^2 + (y-y_c)^2$ and
\begin{equation}
    w(r) = \begin{cases} 
           1 &\mbox{if } r \le 3 \\
           3/r &\mbox{if } r > 3\\
           \end{cases}
    \label{eq:weights}
\end{equation}
Giving equal weight to each annulus stops the large number of pixel values at large radius from dominating the fit, effectively increasing the focus on the cluster at the center.

We break the pixel uncertainty $\sigma(x, y)$ into two components: image-wide sky noise plus Poisson noise from individual sources:
\begin{equation}
    \sigma^2(x, y) = \sigma^2_{\rm sky} + f_d(x, y)
\end{equation}
where $f_d(x, y)$ is the pixel value in electrons, and equals the Poisson variance. To calculate the global sky noise, we use the 3-sigma clipped standard deviation of the pixel values of the entire image.

The model component $f_m(x, y)$ is the convolution of the underlying cluster model with the PSF plus the local background $f_{\rm BG}$, which we assume to be constant over the fitting region. We subsample the pixels for both the model and the empirical PSF, then rebin the resulting model image to the same scale as the data:
\begin{equation}
    f_m(x, y) = \sum_{x_s \in x}\sum_{y_s \in y} ({\rm PSF} \ast \mu)(x_s, y_s)  + f_{\rm BG}
\end{equation}
where $x_s$ and $y_s$ represent subpixel positions that are not integer pixel values like $x$ and $y$, $\ast$ represents convolution, and $\mu$ is the functional form of the fitted profile given by Equation~\ref{eq:mu}.

The last component of Equation~\ref{eq:likelihood} is the prior distribution. We employ a prior on the local background. This is needed because at very low values of $\eta$ (shallow power law slopes), the background can be incorrectly fit by this cluster component rather than a truly flat background. This attributes light to the cluster that should be attributed to the background, incorrectly inflating the enclosed light and therefore $\reff$. Additionally, $\eta$ is strongly degenerate with $a$, so as to give the same value of $\mu$ at some typical radius. Low values of $\eta$ caused by incorrect background fits also lead to unphysically small values for $a$. We find that constraining the background addresses these issues. We first estimate the background and background uncertainty by using sigma clipping to calculate the mean ($\mu_{\rm BG}$) and standard deviation ($\sigma_{\rm BG}$) of all pixels farther than 6 pixels from the cluster center. The mean is used as the mean value of a Gaussian prior on the background. As the background becomes an issue for low values of $\eta$, we condition the width of our prior on it. We use a logistic function that produces a tight prior $\sigma_{\rm prior} = 0.1 \sigma_{\rm BG}$ for low values of $\eta$, while giving a looser constraint $\sigma_{\rm prior} = \sigma_{\rm BG}$ for higher values of $\eta$. This takes the form
\begin{equation}
    \log{P_{\rm prior}} = -\frac{1}{2} \left(\frac{f_{\rm BG} - \mu_{\rm BG}}{\sigma_{\rm prior}}\right)^2
\end{equation}
where 
\begin{equation}
    \sigma_{\rm prior} (\eta) = \sigma_{\rm BG} \left(0.1 + \frac{0.9}{1 + e^{10(1 - \eta)}} \right)
\end{equation}

\begin{figure*}
    \includegraphics[width=\textwidth]{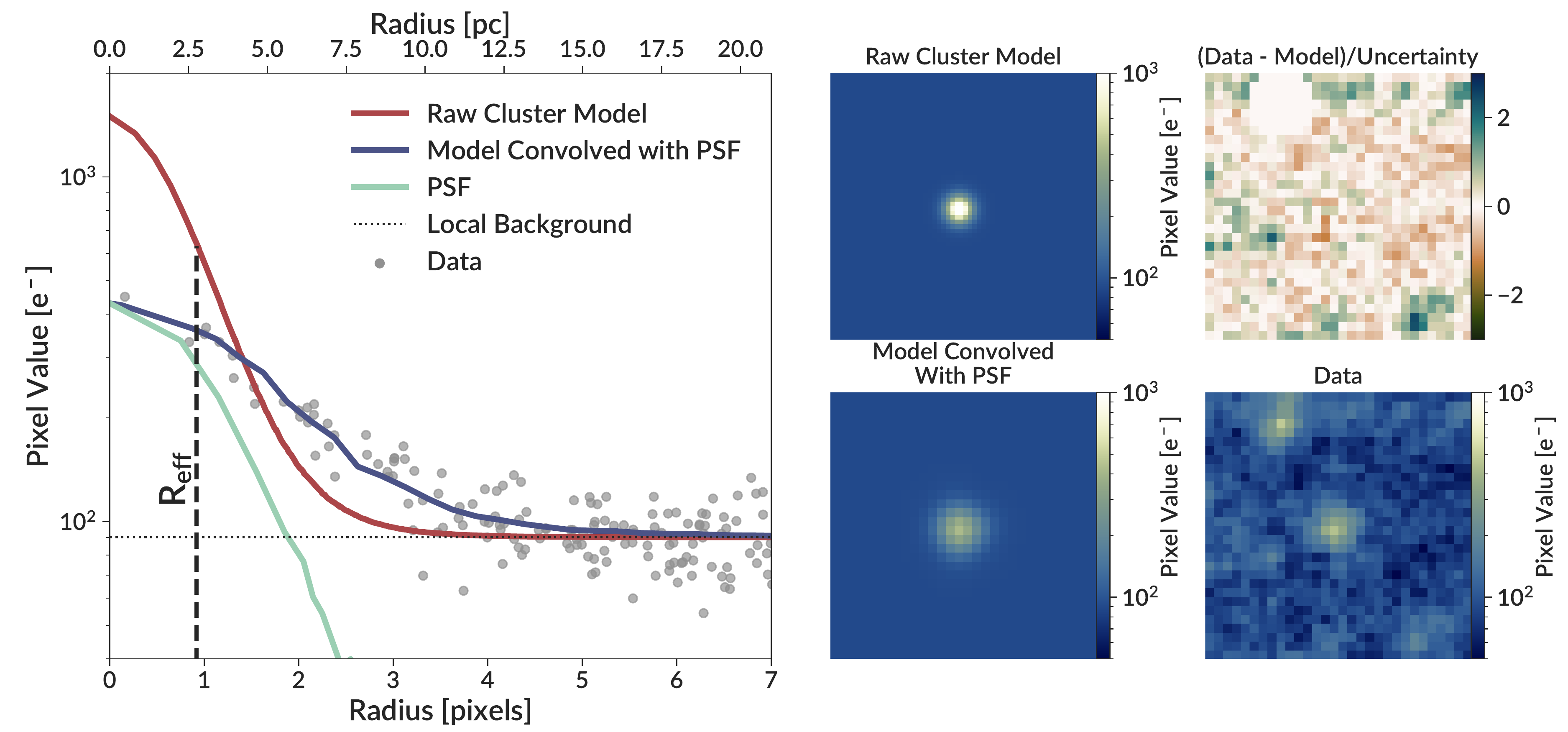}
    \vspace{-3mm}
    \caption{An example of our fitting process. The right set of panels show (in counterclockwise order, starting from the top left) the 2D pixel values for the raw cluster model, the cluster model after being convolved with the PSF, the data, then the residuals after subtracting the PSF-convolved model from the data. Note the masking of the contaminating object in the top right panel. The left panel shows the radial profiles for these components (with the PSF normalization adjusted accordingly). Note that the radial profile is included for illustrative purposes only, the fitting is done using the 2D images.}
    \label{fig:example}
\end{figure*}

The combined effect of using the absolute value of differences, the pixel weights, and the prior on the background is to produce cluster fits that more closely match the cluster itself. They prevent the fit from being drawn towards any surrounding structures, while also allowing the local background to be fit appropriately. This produces trustworthy cluster parameter values. Lastly, to ensure full numerical convergence of the fit, we use multiple sets of initial values for the fit parameters, selecting the final parameter set corresponding to the highest likelihood. An example of a cluster fitted with our method is shown in Figure~\ref{fig:example}.

\begin{figure}
    \includegraphics[width=\columnwidth]{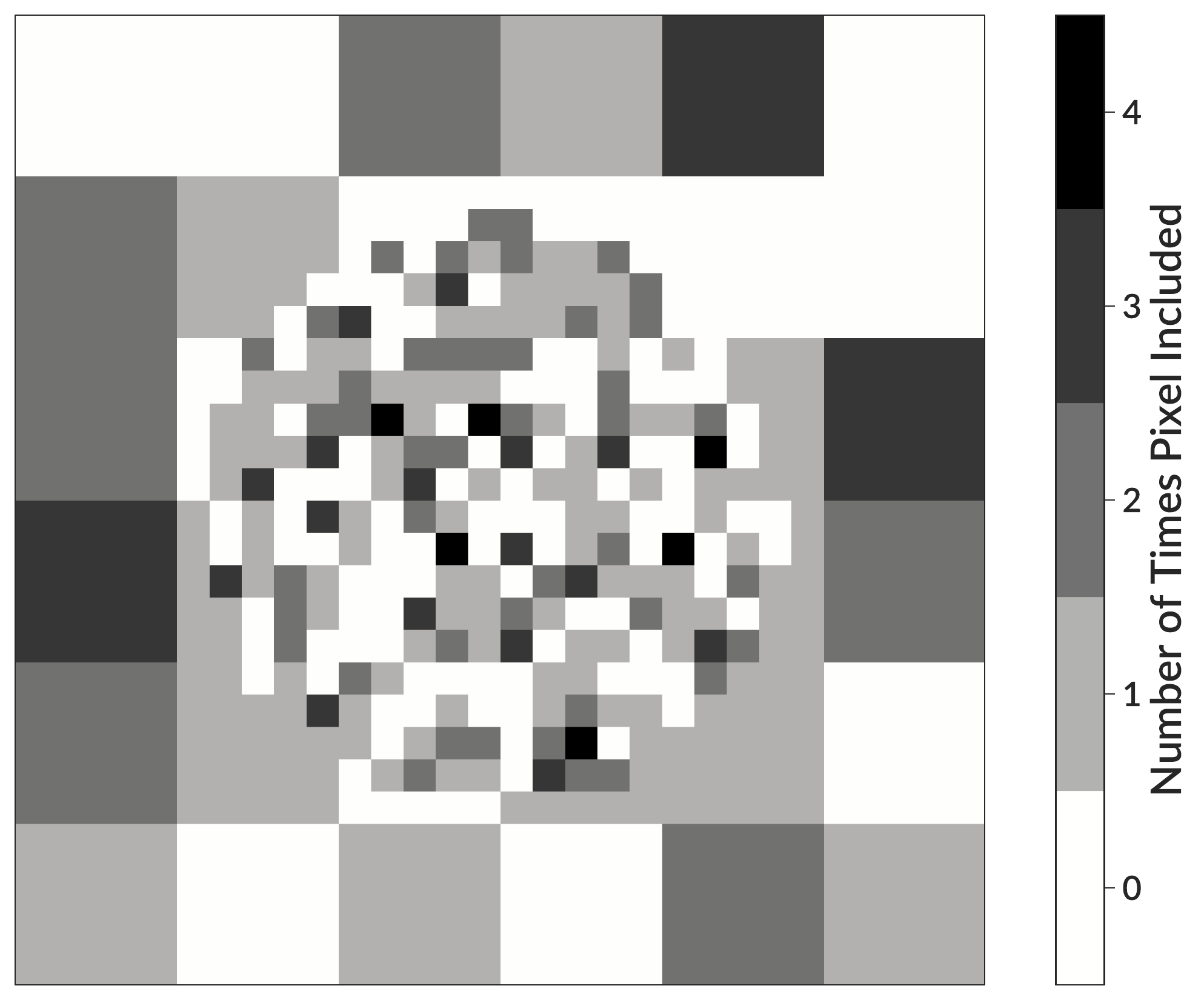}
    \vspace{-3mm}
    \caption{An example of the pixels included in one randomly selected bootstrap iteration. The central region is resampled on a pixel-by-pixel basis, while the outskirts are resampled on 5$\times$5 pixel blocks.}
    \label{fig:bootstrap}
\end{figure}

To determine the distribution of errors of the fit parameters, we perform bootstrapping on the pixels in the snapshot. We split the snapshot into two regions: pixels closer than 9 pixels from the cluster center and pixels outside this region. For the inner region, we resample individual pixels with replacement. For the outer region, we group the pixels into 5x5 blocks, then resample those blocks with replacement. Using blocks in the outskirts does a better job accounting for any missed or faint contaminating sources. If we were to use individual pixels, at least some of the pixels from these sources would be included in a given resampling, while using blocks allow us to exclude these sources completely in certain iterations, giving a better estimate of how these sources affect the cluster fit. Using individual pixels in the center is necessary as the cluster itself may be roughly the size of the 5x5 chunk. Figure~\ref{fig:bootstrap} shows an example of the pixels included in one randomly selected bootstrap realization.

We run the bootstrap realizations until convergence of the fit parameters. Every 20 iterations, we calculate the standard deviation of the distributions of all 8 fit parameters in the accumulated iterations, then compare them to the standard deviations from the last time it was calculated. We stop bootstrapping when the standard deviation of each parameter changes by less than 10 percent. Most clusters required 100--140 iterations to converge. Our reported uncertainties on $\reff$ are marginalized over all other parameters. We use $\reff$ calculated using the original snapshot as the best fit value, then take the 16--84th percentile range of $\reff$ from the bootstrap iterations as the uncertainty.

As we measure everything in pixel values, we need to convert to physical length units. To do this we use the TRGB distances to all LEGUS galaxies provided by \citet{sabbi_etal18}. That work provides independent estimates of distance for each field. For galaxies split between two fields, we use the mean of the two distance estimates for both fields. Lastly, NGC~1566 has an unreliable distance estimate. The TRGB was too faint to be detected in \citet{sabbi_etal18}, and available values in the literature span a wide range (from 6 to 20 Mpc). However, NGC~1566 was identified as being part of the group centered on NGC~1553, which has a measured distance and group radius \citep{kourkchi_tully_17,tully_etal_16}. For NGC~1566, we adopt the distance to NGC~1553 with uncertainty of the group radius.

\subsection{Converting to effective radius}

\begin{figure}
    \includegraphics[width=\columnwidth]{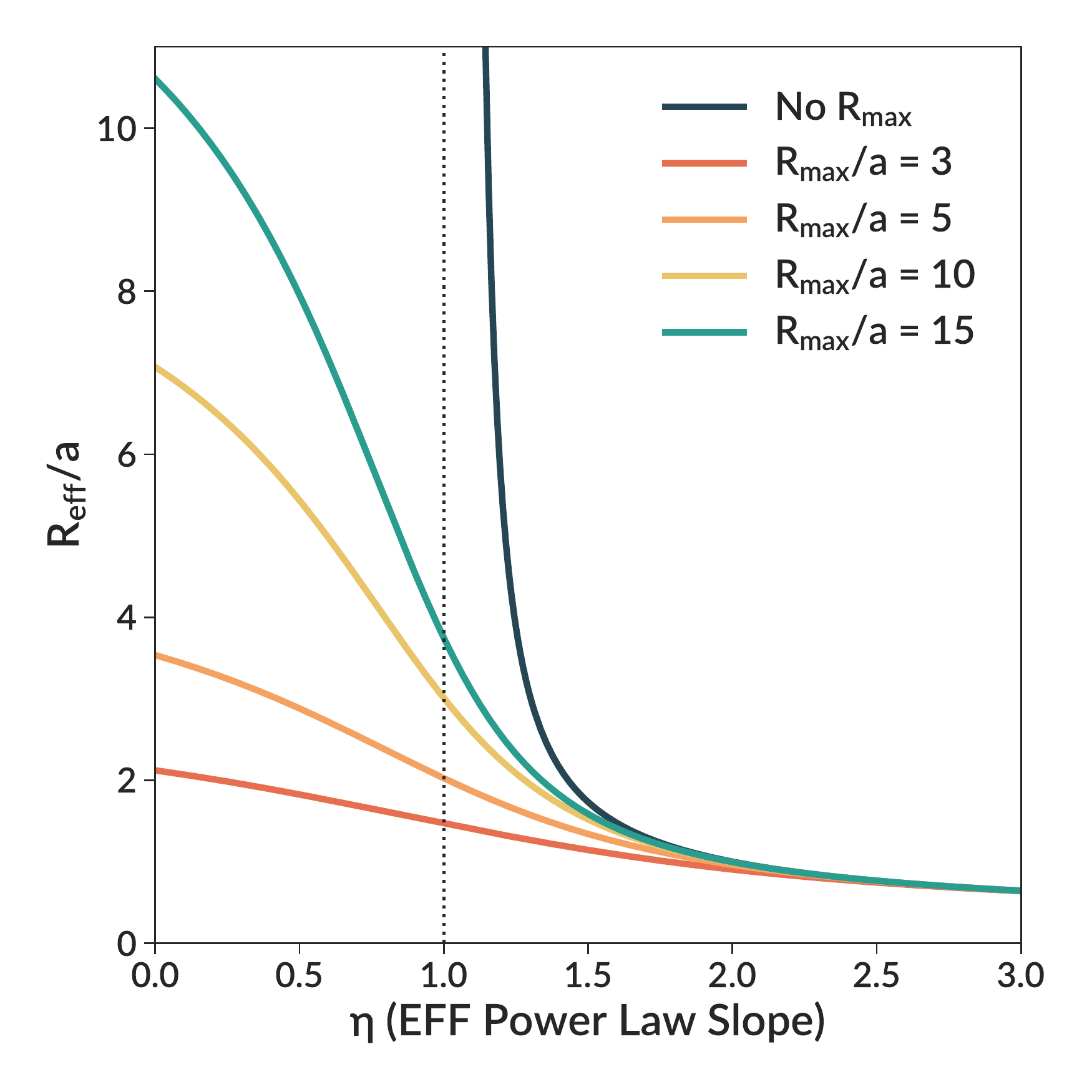}
    \vspace{-7mm}
    \caption{Comparison of the effective radius when calculated with (Eq.~\ref{eq:r_eff_with_rmax}) or without (Eq.~\ref{eq:r_eff_no_rmax}) a maximum radius. We show this for several representative values of $\rmax/a$, as clusters have values across this full range.}
    \label{fig:r_max}
\end{figure}

The effective radius is defined to be the circular radius that contains half of the projected light of the cluster profile. For a circularly symmetric EFF profile, this is
\begin{equation}
    \reff = a \, \sqrt{2^{\frac{1}{\eta-1}} - 1}
    \label{eq:r_eff_no_rmax}
\end{equation}
However, this equation asymptotically approaches infinity as $\eta$ approaches unity, and the total light of the profile is infinite for $\eta \leq 1$. As some cluster fits prefer values of $\eta$ near or below unity, we implement a maximum radius for the cluster profiles, removing this infinity and allowing the effective radius to be well defined for any value of $\eta$. We choose the size of our box (15 pixel radius) as $\rmax$. When using a maximum radius, the effective radius for the circularly symmetric EFF profile is
\begin{equation}
    2 \left[ 1 + \left( \frac{\reff}{a} \right) ^2 \right]^{1 - \eta} = 1 + \left[ 1 + \left( \frac{\rmax}{a} \right)^2 \right]^{1-\eta}.
    \label{eq:r_eff_with_rmax}
\end{equation}
For $\eta \gtrsim 1.5$, this agrees well with Equation~\ref{eq:r_eff_no_rmax}, as shown in Figure~\ref{fig:r_max}. 

\begin{figure}
    \includegraphics[width=\columnwidth]{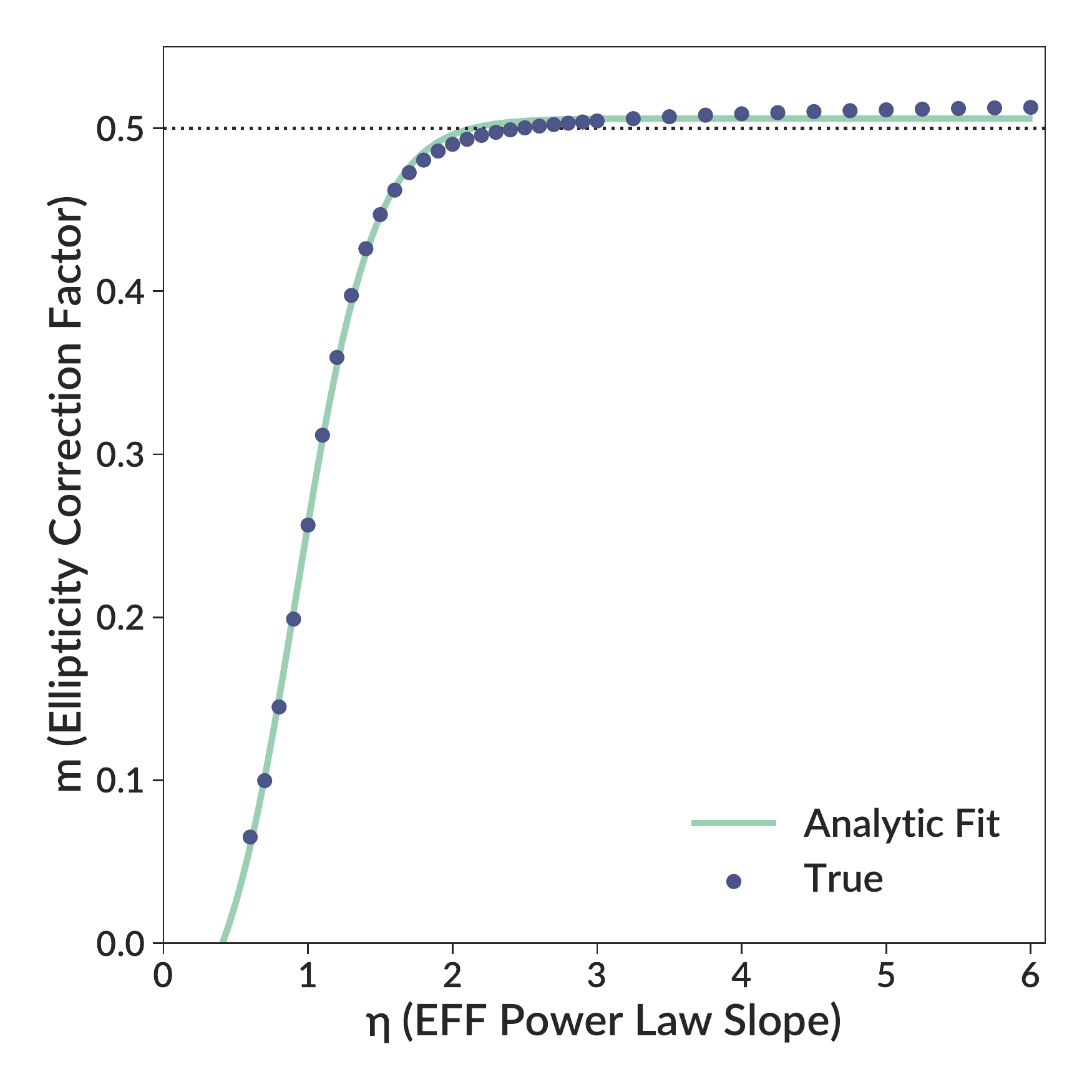}
    \vspace{-7mm}
    \caption{Numerical calculations of the ellipticity correction factor $m$ (see Equation~\ref{eq:ellipticity_k_def}), along with the analytic fit (Equation~\ref{eq:full_ellipticity_correction}). The horizontal dotted line shows $m=0.5$, as used in Equation~\ref{eq:q_correction_simple}.}
    \label{fig:ellipticity}
\end{figure}

A correction is required for an elliptical profile. We empirically determine this correction for the EFF profile as a function of $\eta$ and $q$, by performing numerical integration of elliptical EFF profiles. We determine the circular aperture that contains half of the total light. We use a circular maximum radius $\rmax/a=10$ for this integration, although we find that the results do not depend strongly on the chosen $\rmax$. At a given $\eta$, we find that the relation between true effective radius and the effective radius calculated assuming a circularly symmetric profile (Equation~\ref{eq:r_eff_with_rmax}) is linear with $q$, so we parametrize the correction as 
\begin{equation}
    \frac{R_{\rm eff, true}}{R_{\rm eff, circ}} = 1 + m (q - 1)
    \label{eq:ellipticity_k_def}
\end{equation}
where $m$ is a function of $\eta$. With $m=0.5$, we obtain the commonly used correction 
\begin{equation}
    \frac{R_{\rm eff, true}}{R_{\rm eff, circ}} = 0.5 (1 + q)
    \label{eq:q_correction_simple}
\end{equation}
as found in the ISHAPE manual \citep{larsen_99}. We measure $m$ as a function of $\eta$, and find it is well fit by a logistic-type function, with an RMS deviation of only 0.0043. This gives a final correction of the form:
\begin{equation}
    \frac{R_{\rm eff, true}}{R_{\rm eff, circ}} = 1 + \left(\frac{0.579}{1 + \exp{\left(\frac{0.924 - \eta }{0.266}\right)}} - 0.073\right)(q - 1)
    \label{eq:full_ellipticity_correction}
\end{equation}
The results of the numerical integration and the fit are shown in Figure~\ref{fig:ellipticity}.

\subsection{Cluster fit quality}
\label{sec:quality}

While our fitting procedure is designed to be robust, it does not perform perfectly on all clusters. We exclude clusters with unrealistic parameter values, which we define as a scale radius $a < 0.1$ pixels, $a > 15$ pixels, or an axis ratio $q < 0.3$. We also exclude any clusters where the fitted center is more than 2 pixels away from the central pixel identified by LEGUS. This eliminates 6.7\% percent of the sample.

Additionally, to quantitatively evaluate which clusters have poor fits, we implement a quality metric based on a comparison of the cumulative light profiles of the cluster data and the model (after subtracting the best-fit background from both the data and model). As our primary goal is to evaluate the reliability of $\reff$, the cumulative profile is a strong indicator as it probes all light enclosed within a given radius.

Specifically, our metric uses the cumulative light profile to estimate the half-light radius of the cluster non-parametrically, then compares the enclosed light of the model and data within this radius. The relative difference is
\begin{equation}
    d = \left| \frac{F_{\rm model}(<R_{1/2}) - F_{\rm data}(<R_{1/2}) }{F_{\rm data}(<R_{1/2})} \right|
\end{equation}
where the non-parametric radius $R_{1/2}$ is defined by
\begin{equation}
    F_{\rm data}(<R_{1/2}) = 0.5 F_{\rm data}(<15 {\rm \ pixels}).
\end{equation}
Here $F(<R)$ is the cumulative flux enclosed within a circular radius $R$. We use 15 pixels as the maximum radius as it is the radius of the individual cluster snapshots. 

\begin{figure}
    \includegraphics[width=\columnwidth]{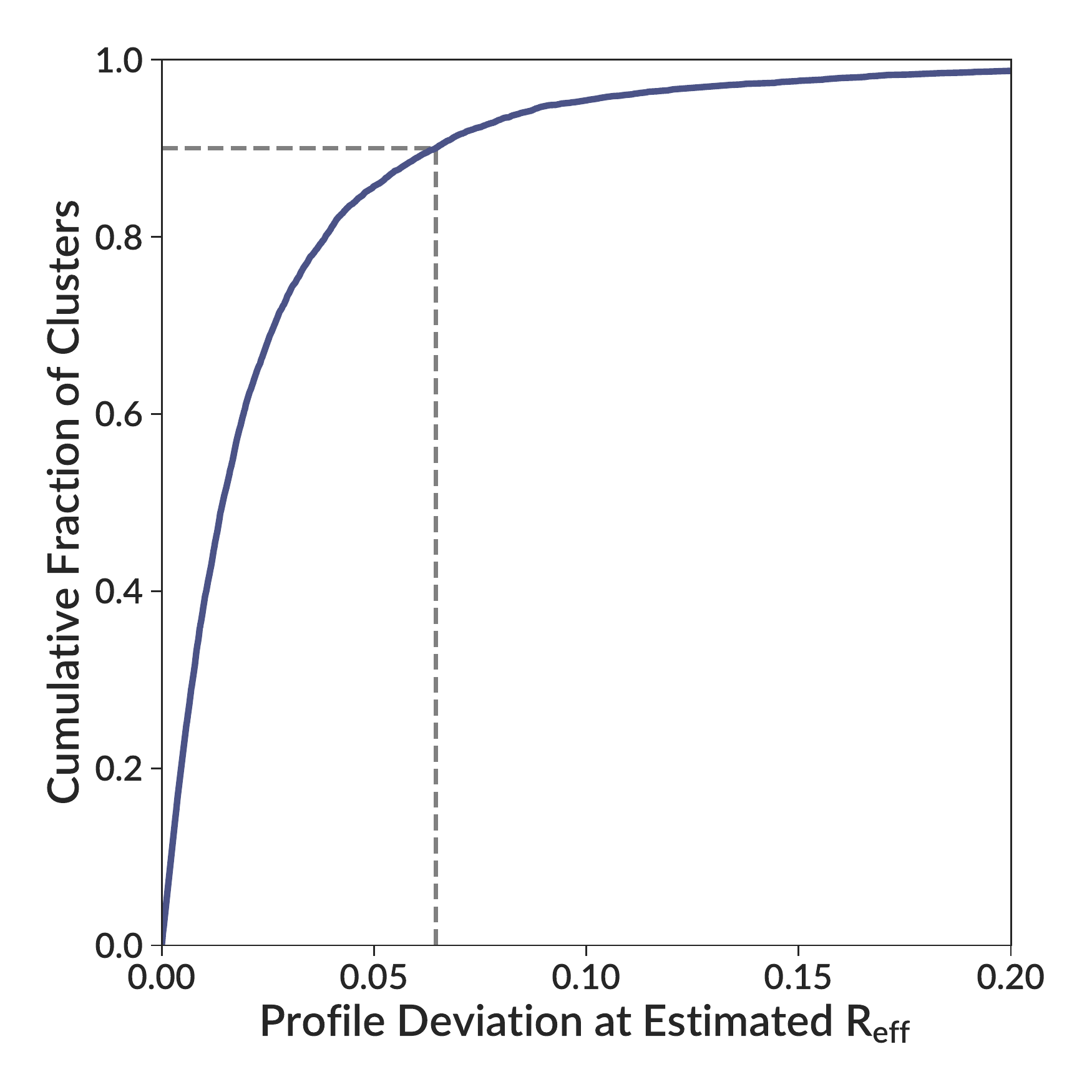}
    \vspace{-7mm}
    \caption{Cumulative distribution for the deviation in the cumulative light profile as described in Section~\ref{sec:quality}. Our analysis excludes clusters that fall above the 90th percentile of this metric, corresponding to approximately 6.5\% deviation of the integrated light within non-parametric $R_{1/2}$.}
    \label{fig:quality_metrics}
\end{figure}

We then calculate the distribution of this metric $d$ for clusters that pass the cuts mentioned at the first paragraph of this section, shown in Figure~\ref{fig:quality_metrics}. The knee of the cumulative distribution is at approximately the 90th percentile, so we use that percentile as our cut. Any cluster above the 90th percentile will be excluded from the analysis in the rest of this paper. This results in a final sample of 6097 clusters with reliable radius measurements. This 90th percentile cut corresponds to about 6.5\% error on the light enclosed within the estimated effective radius, indicating the high quality of the fits we keep.

\subsection{Artificial cluster tests}

To test the ability of the pipeline to recover the effective radius of small clusters, we perform artificial cluster tests. We generate 150 synthetic clusters following the EFF profile. These clusters have magnitudes from 20 to 24, $1.25 \leq \eta \leq 2.5$, and $0.03 \leq \reff \leq 3$ pixels. All parameters are uniformly distributed within these ranges. To calculate the cluster magnitude, we follow the LEGUS pipeline as described in \citet{Adamo_etal_17}. We use a circular aperture with a radius of 4 pixels (chosen as the integer pixel value that contains 50\% of the flux of typical clusters) and a local sky annulus located at 7 pixels with a width of 1 pixel, then apply the average aperture corrections for the NGC~628-c field. The artificial clusters span the range of the magnitudes of real clusters in this field.

\begin{figure}
    \includegraphics[width=\columnwidth]{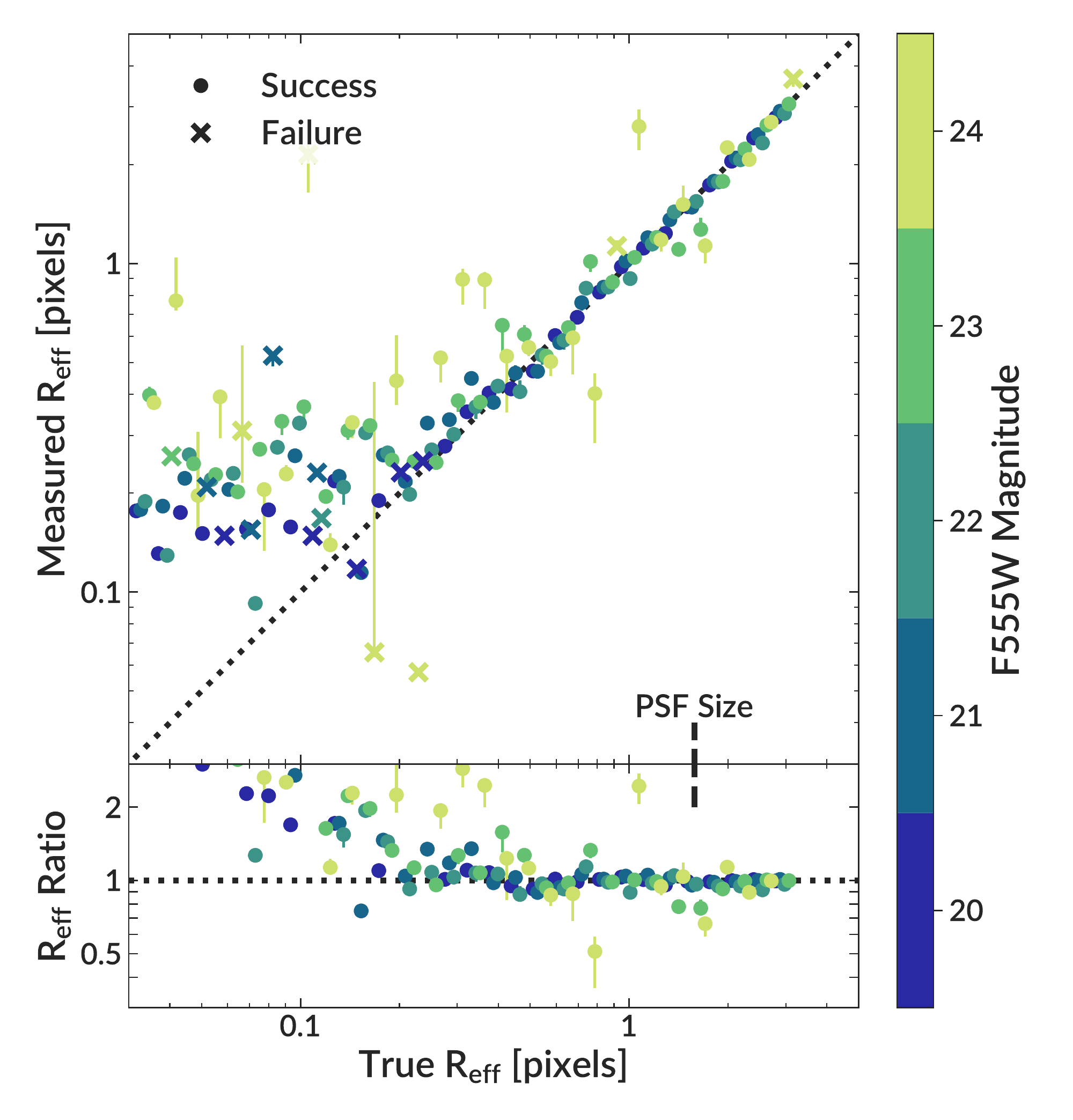}
    \vspace{-7mm}
    \caption{Results of artificial cluster tests. The top panel shows a comparison of the true effective radius to that measured by our pipeline. Solid circles show fits that the pipeline identified as successful, while the crosses show failures. The bottom panel shows the ratio of the measured effective radius to the true effective radius. Only successful fits are shown in this bottom panel. The dashed line spanning both panels indicates the PSF size in pixels for the NGC~628-c field, which is the field into which clusters were inserted.}
    \label{fig:artificial_clusters}
\end{figure}

We convolve these models with the PSF for the NGC~628-c field, add Poisson noise, and insert these artificial clusters into the NGC~628-c field. We then run our pipeline on this new image to measure their effective radii. The results of this test are shown in Figure~\ref{fig:artificial_clusters}.

The pipeline is able to accurately measure cluster radii down to about 0.3 pixels. Below this point, the pipeline systematically overestimates the true radius. The performance of the pipeline does depend on magnitude, as faint clusters with magnitude 24 (the limit of the clusters in the LEGUS catalog for NGC~628-c) have a much wider dispersion than brighter clusters, even at larger radii. A visual examination of these fits shows that contamination and noise are the primary causes of this dispersion. Faint sources rise above the background less than bright sources, so variations in the background can influence the fit more. Additionally, the Poisson pixel noise of the source itself can influence the fit, even for artificial clusters placed in a region where the background is smooth. Due to the nature of Poisson noise, this affects faint clusters the most.

Figure~\ref{fig:artificial_clusters} also shows the ability of the pipeline to detect when clusters have poor fits. Many of the catastrophic failures are correctly identified as failures. However, for the very compact clusters the pipeline identifies many fits that actually overestimate $\reff$ as reliable. This is likely because when $\reff$ is much smaller than the PSF, the observed cluster is not very different from the PSF. A slightly larger model is still very similar to the data, making the pipeline identify the fit as a success. This also means that generally the error bars may be underestimated for clusters with $\reff < 0.3$ pixels. Nevertheless, for most clusters our pipeline gives reliable measurements. Importantly, if $\reff$ is measured to be small, then it really is small. These trends are true for all values $\eta \gtrsim 1.25$. For values of $\eta \sim 1$, we found larger scatter.

\section{Results}
\label{sec:results}
\subsection{Comparison to Ryon et al. (2017)}

\begin{figure*}
    \includegraphics[width=\textwidth]{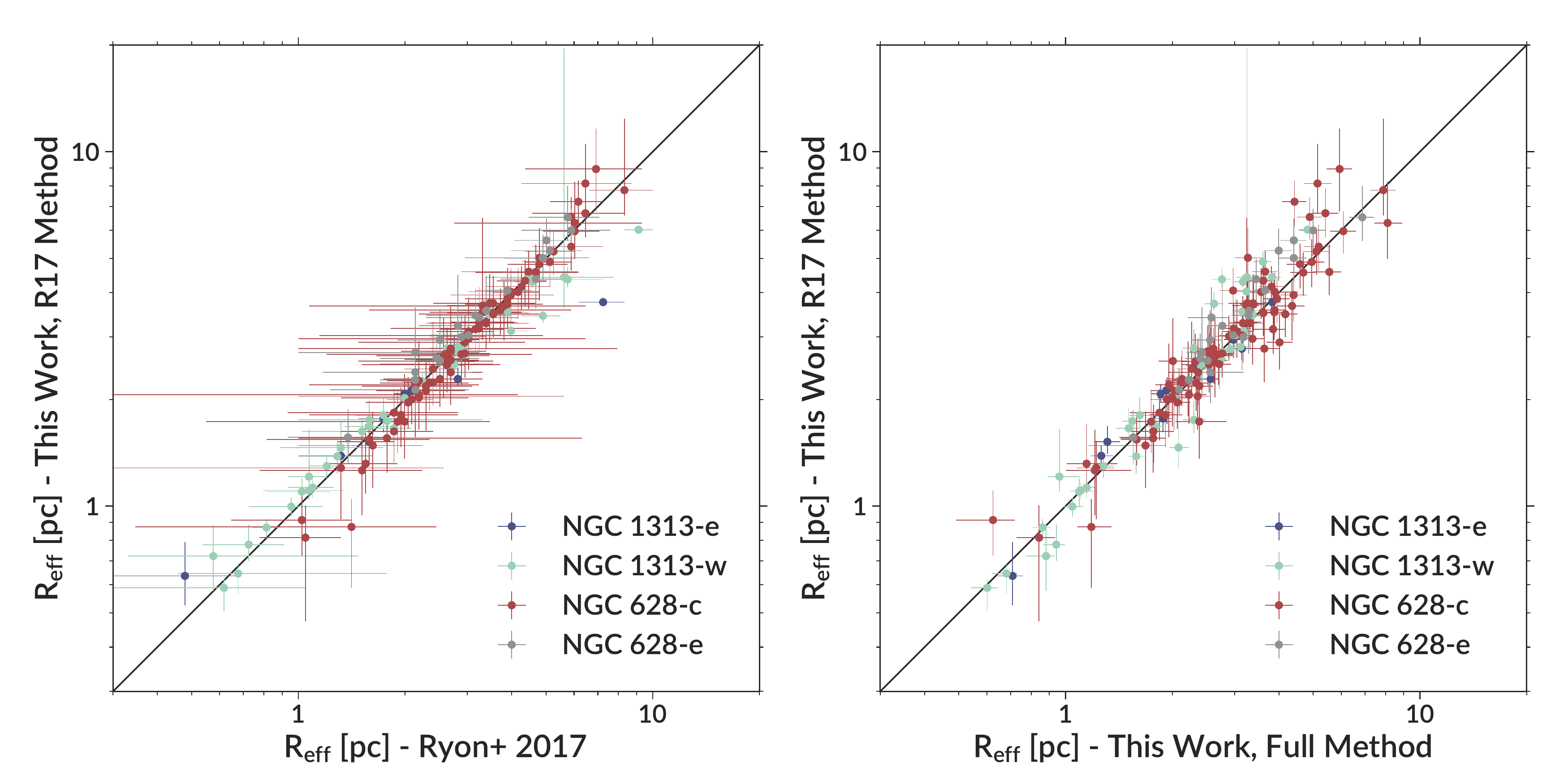}
    \vspace{-5mm}
    \caption{Comparison of our cluster effective radii using different fitting methods for NGC~628 and NGC~1313. The left panel shows a comparison to those of \citet{ryon_etal_17} using the same fitting method. The right panel shows a comparison of this method to our full method used in the rest of the paper.}
    \label{fig:ryon_comparison}
\end{figure*}

\citetalias{ryon_etal_17} used \galfit\ to measure the effective radii of clusters in NGC~1313 and NGC~628, two of the galaxies in the LEGUS sample. To validate our method, we compare our measurements to those of \citetalias{ryon_etal_17}. When making this comparison, we perform a separate round of fitting with several modifications to our method to match what was done in \citetalias{ryon_etal_17}. We do not mask any contaminating sources, do not use radial weighting, and use the square of differences rather than the absolute value. When postprocessing these results, we do not use a maximum radius to calculate the effective radii (instead using Equation~\ref{eq:r_eff_no_rmax}), we use the simple ellipticity correction (Equation~\ref{eq:q_correction_simple}), and we use the same distances to NGC~1313 and NGC~628 as \citetalias{ryon_etal_17} did \citep{jacobs_etal09,olivares_etal10}. These changes ensure consistency with the \citetalias{ryon_etal_17} method.

The left panel of Figure~\ref{fig:ryon_comparison} shows the results of that comparison. Following \citetalias{ryon_etal_17}, only clusters with $\eta>1.3$ are shown in this plot, as $\reff$ given by Equation~(\ref{eq:r_eff_no_rmax}) is unreliable for lower values. To quantify the deviation, we use the RMS error, defined as 
\begin{equation}
    \mathrm{RMS} = \sqrt{\frac{1}{N} \sum \frac{ \left( R_{\mathrm{eff, R17}} - \reff \right)^2}{\sigma_{\mathrm{R17}}^2 + \sigma^2}}
\end{equation}
where $\sigma$ and $\sigma_{\mathrm{R17}}$ are the error on $\reff$ in this work and \citetalias{ryon_etal_17}, respectively. This RMS deviation is 0.55, indicating excellent agreement.

In the right panel of Figure~\ref{fig:ryon_comparison} we show a comparison of our full method to our \citetalias{ryon_etal_17}-like method. These two methods show good agreement across the full radius range, with no significant deviations. The RMS deviation here is 1.62. This higher value is primarily driven by the error bars, which are smaller than those obtained in \citetalias{ryon_etal_17}. In addition, a small number of clusters have significantly different radii between the two methods. The majority of these discrepant clusters have a small value for $\eta$ (typically just above the cutoff of $\eta > 1.3$ for inclusion in this plot), where the use of a maximum radius has the greatest effect on $\reff$ (see Figure~\ref{fig:r_max}).

In the rest of this paper we analyze the results obtained with our full fitting method.

\begin{figure*}
    \includegraphics[width=\textwidth]{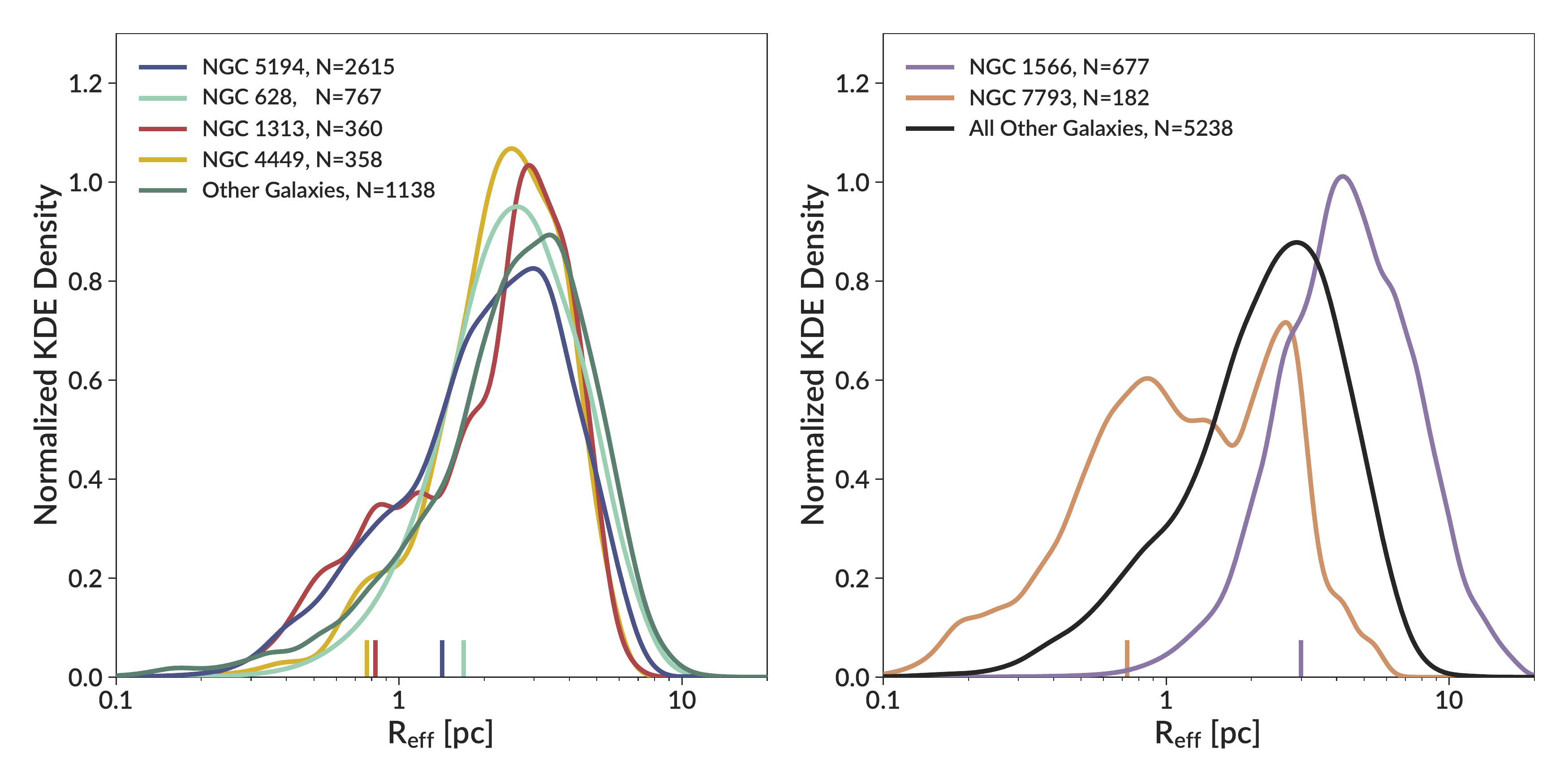}
    \vspace{-5mm}
    \caption{Kernel density estimation of the cluster radius distributions of the galaxies with the most clusters. The line for each galaxy shows the summed Gaussian kernels representing its clusters, where we use a width equal to twice each cluster's radius error. Each curve is normalized to the same area for comparison purposes. The left panel shows galaxies with similar cluster distributions, while the right panel shows two galaxies with different distributions. Note that the ``Other Galaxies'' in the left panel do not include the two discrepant galaxies NGC~1566 and NGC~7793 shown in the right panel, while the ``All Other Galaxies'' in the right panel include all galaxies shown in the left panel. The tick marks at the bottom show the pixel size in parsecs for the images of each galaxy.}
    \label{fig:radius_dist}
\end{figure*}

\subsection{Cluster radius distribution}
\label{sec:radius_dist}

In Figure~\ref{fig:radius_dist} we show the distribution of effective radii measured in the entire LEGUS sample with our full fitting method. For most galaxies the distributions are remarkably similar. They have a peak at $\reff\approx 3$~pc, an extended tail to below 1~pc, and a sharper cutoff at large radii. This peak at $\reff\approx 3$~pc has been seen consistently in other studies of young clusters \citep[e.g.][]{meurer_etal_95,larsen_99,scheepmaker_etal_07,bastian_etal_12,ryon_etal_15,ryon_etal_17,cantat_gaudin_etal_18}. Galaxies with similar distributions are shown in the left panel of Figure~\ref{fig:radius_dist}, while two galaxies NGC~1566 and NGC~7793 with discrepant shapes are shown in the right panel and will be discussed below. While we only show several galaxies individually in the left panel of Figure~\ref{fig:radius_dist} for clarity, an examination of all galaxies shows that they have very similar distributions. In Table~\ref{tab:galaxies} we include the quartiles of the cluster radius distribution of each galaxy as another method of quantifying their distributions.

To characterize this common distribution, we create a stacked distribution of clusters from all galaxies other than the two discrepant. The distribution is shown as the comparison line in the right panel of Figure~\ref{fig:radius_dist} and has a sharp peak at 2.9~pc. We compared it to several common analytical distributions and found that neither normal nor lognormal functions are good fits, due to the asymmetric shape of the observed distribution. Instead we find that the Weibull distribution produces an excellent match:
\begin{equation}
    \frac{dN}{d\reff} = \frac{k}{\lambda} \left( \frac{\reff - R_0}{\lambda} \right)^{k - 1} \; \exp \left[ - \left( \frac{\reff - R_0}{\lambda} \right)^k\right]
    \label{eq:weibull}
\end{equation}
with $k=2.22$, $\lambda = 3.64$~pc, and $R_0 = 0.185$~pc.

To quantitatively test whether the individual galaxy samples are statistically consistent with being drawn from the same distribution, we employ the one-sided Kolmogorov--Smirnov test. We find that of the 31 galaxies, 16 have $p-$value $> 0.01$ (13 with $p-$value $> 0.05$), indicating that they are not inconsistent with the stacked distribution. However, of the 13 galaxies with more than 50 clusters, only 4 have $p > 0.01$ (2 with $p > 0.05$). The large number of clusters in these galaxies provides high statistical significance to formally distinguish the distributions. Still, the individual distributions exhibit strong visual similarity.

Two galaxies show cluster distributions significantly different from the rest. NGC~1566 appear shifted to larger radii than other galaxies. It has less low radius clusters, a peak at larger radii (4.2~pc compared to 2.9~pc for the stacked distribution), and more high radius clusters than any other galaxy. Selection effects may be partly responsible. At the adopted distance of NGC~1566 of 15.6~Mpc, 1 pixel covers 3~pc and our PSF model has an effective radius of 4.3~pc. Small clusters may not be resolved and therefore not included in the LEGUS catalog. While a full characterization of the LEGUS selection effects is beyond the scope of this paper, \citet{Adamo_etal_17} examined the completeness as a function of cluster radius in NGC~628 at a distance of 8.8~Mpc. The concentration index cut excluded roughly 50\% of clusters with $\reff=1$~pc. This selection effect is not significant for most galaxies, as the peak of the cluster radius distribution is at higher radii, and most galaxies are closer than NGC~628. But since NGC~1566 is at approximately twice the distance of NGC~628, we can expect that its observations will be incomplete below 2~pc. This could explain the dearth of small clusters in NGC~1566, but would not explain the overabundance of large clusters. Our adopted distance could be responsible for this. As mentioned at the end of Section~\ref{sec:fitting}, the distance to NGC~1566 is uncertain, with distance estimates ranging from 6 to 20~Mpc. If our adopted distance of 15.6~Mpc is an overestimate, our cluster radius measurements will also be overestimated. Adopting a distance of 11~Mpc rather than 15.6~Mpc would bring it in line with the distributions of other galaxies, thus effectively treating this distribution like a standard ruler \citep{jordan_etal_05}. Future distance measurements may be able to resolve this and determine whether the cluster radius distribution in NGC~1566 is significantly different than that of other galaxies. 

The other discrepant galaxy, NGC~7793, has a double peaked distribution that is much broader than that in other galaxies. One peak is near the 3 pc peak seen in other galaxies, while another is at $\sim 0.8$ pc. The reason for this is unclear. While NGC~7793 is split into two fields, both fields show the same double peaked distribution. Its specific star formation rate is within the range of other galaxies. It is closer than most other galaxies in the sample, meaning the smallest clusters are more likely to be included, but other galaxies with similar distances do not show this bimodal distribution. A visual examination of the spatial distribution within NGC~7793 of the clusters belonging to each peak does not show any striking trends. An examination of the age and mass of the clusters shows that, compared to other galaxies, NGC~7793 has more small young clusters and no large young clusters. Additionally, the age distribution is bimodal, with a deficit of intermediate age clusters.  Roughly speaking, this results in the low-radius peak being mostly young, low-mass clusters, while the high-radius peak is mostly old, high-mass clusters. Future detailed studies of NGC~7793 may be needed to understand its cluster population in more detail.

\subsection{Cluster mass-radius relation for all LEGUS galaxies}
\label{sec:mass_radius}

\begin{figure*}
    \includegraphics[width=\textwidth]{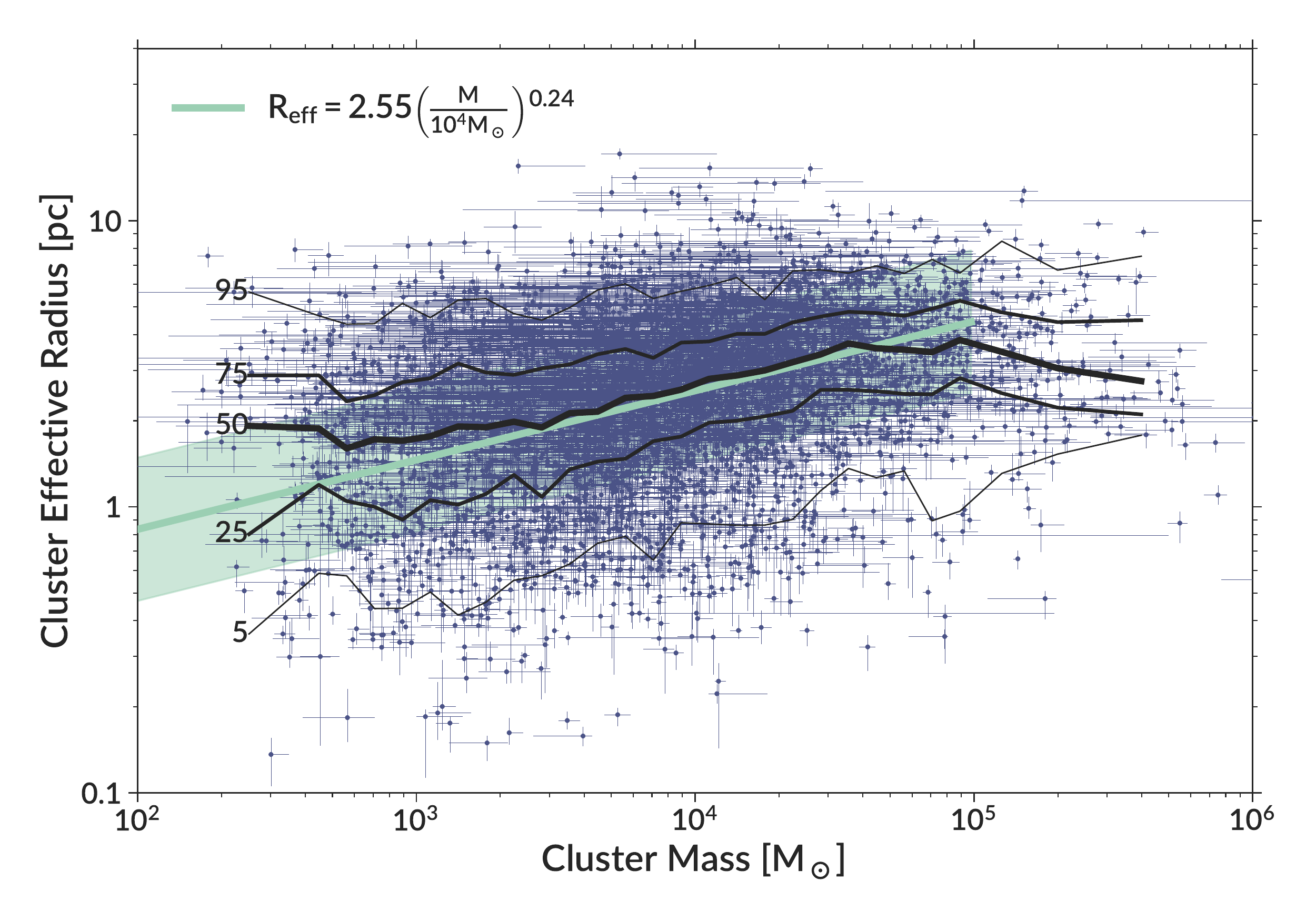}
    \vspace{-9mm}
    \caption{The mass-radius relation for the clusters in LEGUS. Clusters of all ages are included in this plot. The black lines show then 5, 25, 50, 75, and 95th percentiles of radii at a given mass. The solid line shows the best fit linear relation, with the shaded region showing the intrinsic scatter.}
    \label{fig:mass_radius}
\end{figure*}

Figure~\ref{fig:mass_radius} shows the mass-radius relation for the clusters in our sample. A clear mass-radius relation is visible, albeit with a shallow slope. To guide the eye, lines indicate the 5, 25, 50, 75, and 95th percentiles of the radius at a given mass. 

This plot shows the full mass range of the LEGUS clusters. However, masses below 5000~$\Msun$ measured by the deterministic method used in LEGUS may be unreliable, as the assumption of a fully-sampled IMF is no longer valid \citep{maiz_apellaniz_09,silvavilla_larsen_11,krumholz_etal_15}. To account for this, previous work with the LEGUS data, such as \citet{Adamo_etal_17} and \citetalias{ryon_etal_17}, restricted to clusters with masses above 5000~$\Msun$. As LEGUS is not complete for 5000~$\Msun$ clusters at old ages, these papers selected clusters younger than 200~Myr. This produces a complete sample using clusters with the most reliable masses. We take a different approach in this paper, and use the full mass and age ranges of clusters with good SED fits, adjusting their errors to account for the systematic errors in the SED fitting. If we were to consider only clusters more massive than 5000~$\Msun$, it would exclude about a third of the complete sample and greatly decrease the dynamic range of the mass-radius relation. Section~\ref{sec:selection} below discusses possible effects of incompleteness.

As discussed in \citet{krumholz_etal_15}, clusters have poor fits -- as quantified by the $Q(\chi^2)$ statistic --  when the assumption of a fully-sampled IMF is violated. To avoid these unreliable low-mass clusters, we restrict ourselves to clusters with $Q(\chi^2) > 10^{-3}$. Of the 6097 clusters with successfully measured radii, 5105 pass this further cut. In addition to low-mass clusters, we also find qualitatively that this cut removes many clusters with high masses ($>10^5 \Msun$) and very small radii ($\reff < 1$ pc) that were outliers from the mass-radius relation due to their unreliable mass. 

In addition, \citet{krumholz_etal_15} show that the deterministic LEGUS mass uncertainties are likely underestimated for low-mass clusters. To correct for this, we compare the mass uncertainty in \citet{krumholz_etal_15} to the uncertainty in the LEGUS catalogs. For clusters below 5000 $\Msun$, the median difference in uncertainty is 0.16 dex. We add this 0.16 dex correction to the mass uncertainty of all clusters below 5000 $\Msun$ when performing our fits.

This produces a sample across the full mass range that includes only the most reliable low-mass clusters and adjust their errors to account for the systematic error in the deterministic LEGUS SED fitting. Our resulting sample may not be complete (as we will be missing old, low-mass clusters), but we will discuss these selection effects throughout the rest of the paper.

We fit this mass-radius distribution assuming a power law relation:
\begin{equation}
    \hat{R}_\mathrm{eff}(M) = R_4 \left(\frac{M}{10^4 \Msun} \right) ^\beta
    \label{eq:mass_radius_relation}
\end{equation}
such that the normalizing factor $R_4$ is the effective radius at $10^4 \Msun$. We incorporate errors in both mass and radius and account for the intrinsic scatter by minimizing projected displacements of data points from the fit line as outlined in \citet{hogg_etal_10}. In Appendix \ref{appendix:fitting} we describe this in more detail and compare it to a hierarchical Bayesian model that includes a treatment of selection effects. We found that neither the fitting method nor the inclusion of selection effects made a difference in the fit parameters, so we use this simpler method. We restrict the fitting to clusters below $10^5 \Msun$, as the relation appears to flatten above that mass, possibly because of small-number statistics. We determine errors on the fit via bootstrapping. For this full sample, our best fit power law slope is $\beta=0.24$, with an intrinsic scatter of 0.25 dex. Restricting to clusters with ages younger than 1 Gyr produces the same result, primarily because most clusters in LEGUS are younger than 1~Gyr.

With this large sample, we can investigate the cluster mass-radius relation for various subsamples of the data. For all cases below, we restrict our sample to clusters younger than 1 Gyr and less massive than $10^5 \Msun$. Table~\ref{tab:fits} shows the fitting parameters for all the subsets discussed below.

\begin{figure}
    \includegraphics[width=\columnwidth]{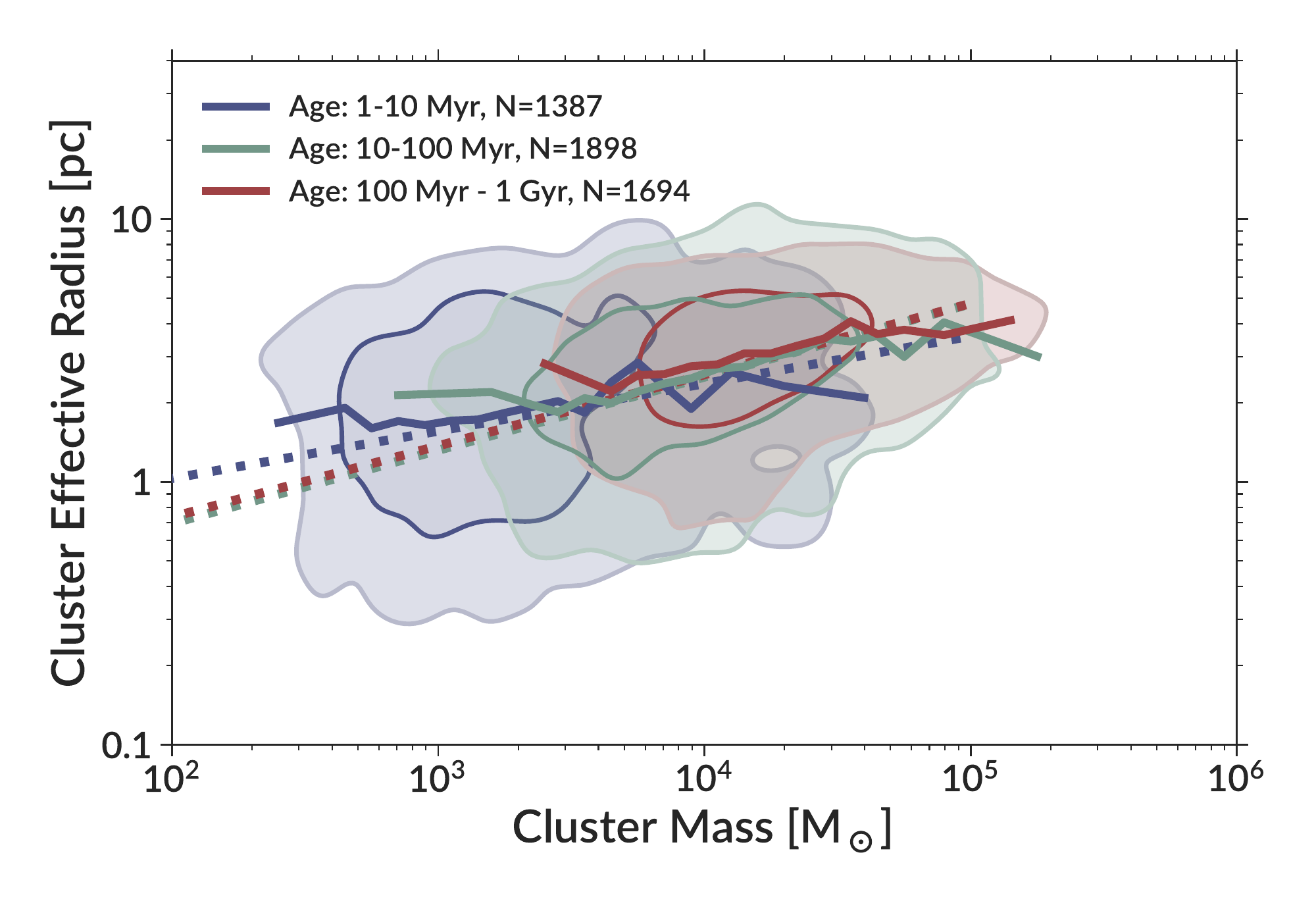}
    \vspace{-7mm}
    \caption{The mass-radius relation for the clusters in LEGUS, split by age. Contours enclose 50 and 90\% of the clusters in each bin, and are smoothed by a kernel of 0.08 dex. The dashed lines show the fits, while the solid lines show the running median in each age bin.}
    \label{fig:mass_radius_agesplit}
\end{figure}

\begin{table*}
    \centering
    \caption{Fits to the cluster mass-radius relation for various subsets of the data. All fits were done using clusters with masses below $10^5\Msun$, while all fits other than the first row were done using clusters younger than 1 Gyr. Note that the fits including other datasets are sensitive to how these datasets are weighted, adding systematic uncertainties to the fit parameters.}
	\begin{tabular}{llcccc}
		\toprule
		Selection & $N$ & $\beta$: Slope & $R_4$: $\reff$(pc) at $10^4\Msun$ & Intrinsic Scatter & $\log{M}$ percentiles: 1--99 \\ 
		\midrule
		Full LEGUS Sample & 5105 & 0.242 $\pm$ 0.010 & 2.548 $\pm$ 0.022 & 0.250 $\pm$ 0.003 & 2.57 -- 5.40 \\ 
		1 Myr -- 1 Gyr & 4979 & 0.246 $\pm$ 0.010 & 2.554 $\pm$ 0.022 & 0.250 $\pm$ 0.003 & 2.57 -- 5.26 \\ 
		\midrule
		Age: 1--10 Myr & 1387 & 0.180 $\pm$ 0.028 & 2.365 $\pm$ 0.106 & 0.319 $\pm$ 0.006 & 2.40 -- 4.89 \\ 
		Age: 10--100 Myr & 1898 & 0.279 $\pm$ 0.021 & 2.506 $\pm$ 0.035 & 0.238 $\pm$ 0.005 & 2.91 -- 5.24 \\ 
		Age: 100 Myr -- 1 Gyr & 1694 & 0.271 $\pm$ 0.027 & 2.558 $\pm$ 0.048 & 0.198 $\pm$ 0.005 & 3.46 -- 5.40 \\ 
		\midrule
		LEGUS + MW & 6158 & 0.296 $\pm$ 0.002 & 2.555 $\pm$ 0.022 & 0.225 $\pm$ 0.003 & 0.93 -- 5.22 \\ 
		LEGUS + External Galaxies & 5874 & 0.229 $\pm$ 0.008 & 2.561 $\pm$ 0.020 & 0.244 $\pm$ 0.003 & 2.36 -- 5.31 \\ 
		LEGUS + MW + External Galaxies & 7053 & 0.292 $\pm$ 0.002 & 2.567 $\pm$ 0.020 & 0.222 $\pm$ 0.003 & 0.97 -- 5.26 \\ 
		\bottomrule
	\end{tabular}
    \label{tab:fits}
\end{table*}

Figure~\ref{fig:mass_radius_agesplit} shows the mass-radius relation split by cluster age. The mass range of the three bins is clearly different, due to LEGUS's absolute V band magnitude cut. Evolutionary fading results in only massive clusters being detected at older ages. To demonstrate this, we include the mass range spanned by the 1--99th percentile of each sample in Table~\ref{tab:fits}. In all 3 bins, we detect a significantly nonzero slope of the mass-radius relation. The running medians of each panel are quite similar, especially for the two oldest age bins, which deviate from the 1--10 Myr bin at $M > 10^4 \Msun$. This matches what we find in the formal fit, where the two older bins have a slope and normalization indistinguishable from each other, while the 1--10 Myr bin has a significantly shallower slope. In addition, the intrinsic scatter decreases with age. 

\begin{figure*}
    \includegraphics[width=\textwidth]{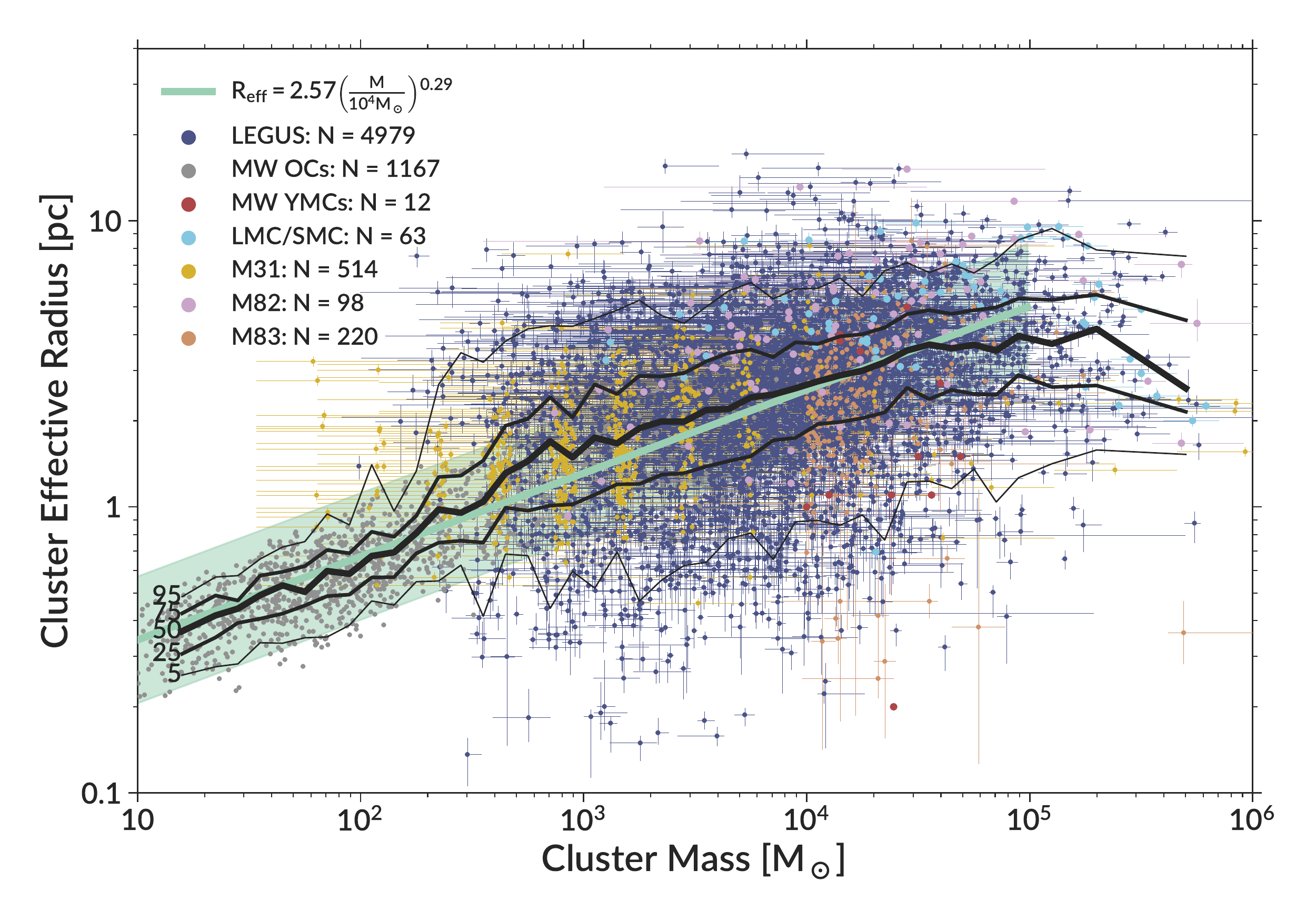}
    \vspace{-8mm}
    \caption{The mass-radius relation for the clusters in LEGUS as well as external datasets, as described in the text. Only clusters with ages less than 1 Gyr are shown here. The black lines show then 5, 25, 50, 75, and 95th percentiles of radii at a given mass. Note that a small random shift has been applied to the original discrete M31 masses for visual purposes.}
    \label{fig:mass_radius_other_samples}
\end{figure*}

Additionally, we supplement the LEGUS sample with several other large samples for young star clusters, mostly from the compilation of \citet{krumholz_etal_19_review}. In all the samples below, we restrict to clusters with an age less than 1 Gyr and masses below $10^5 \Msun$, as done for our main fits. This age cut means that we do not include any globular clusters. Additionally, some of the samples in \citet{krumholz_etal_19_review} are for galaxies already included in this paper (namely NGC~628, NGC~1313, and NGC~5194), so we do not include them again here.

We include Milky Way open clusters (OCs) within 2 kpc of the Sun from \citet{kharchenko_etal_13}, who measured King parameters for these clusters. Following \citet{krumholz_etal_19_review}, we calculate mass using Equation 3 of \citet{piskunov_etal_07}, Oort constants from \citet{bovy_17}, and the distance from the Sun to the Galactic center from \citet{bland-hawthorn_gerhard_16}. We also include the sample of 12 Milky Way YMCs compiled in \citet{krumholz_etal_19_review}.

We additionally include samples from several external galaxies. \citet{mackey_gilmore_03a,mackey_gilmore_03b} measured radii for 53 clusters in the LMC and 10 clusters in the SMC. EFF profiles were fit to the surface brightness profiles of clusters. These surface brightness profiles were also used to obtain the total luminosity of each clusters, which was converted into the cluster mass by using mass-to-light ratios.

The Panchromatic Hubble Andromeda Treasury (PHAT) survey identified stellar clusters in M31 \citep{johnson_etal_12,fouesneau_etal_14}. Half-light radii were determined by interpolating the flux profile to find the radius in arcseconds that includes half of the light \citep{johnson_etal_12}. We then use a distance of 731 kpc to M31 \citep{wagner_kaiser_etal_15} to convert the radii to parsecs. Masses were determined using a Bayesian SED fitting method that explicitly accounts for the stochastic sampling of the IMF \citep{fouesneau_etal_14}. 

In a series of papers, \citet{cuevasotahola_etal_20} and \citet{cuevasotahola_etal_21} calculated structural parameters for 99 star clusters in M82. They fit EFF, King, and Wilson profiles to the surface brightness profiles, finding that the EFF profile best represents the clusters in their sample. Similarly to \citet{mackey_gilmore_03a,mackey_gilmore_03b}, masses are determined by applying a mass-to-light ratio to fitted luminosities.

We also include the clusters in M83 measured by \citet{ryon_etal_15}. Radii are measured by fitting an EFF profile to the 2D light profile using \galfit, as in \citetalias{ryon_etal_17}. Masses are derived in \citet{silvavilla_etal_14} and are done using an SED-fitting method similar to that used in LEGUS \citep{adamo_etal_10}. 


Figure~\ref{fig:mass_radius_other_samples} shows the mass-radius relation including these data sources. In the fit shown in this figure, we give each cluster equal weight, no matter which dataset it comes from. 
The addition of the MW OCs in particular extends the mass-radius relation down to very small masses, while the other samples overlap nicely with LEGUS. Including the MW clusters produces a steeper slope than the fit using only LEGUS clusters ($\beta = 0.296$). In addition, the intrinsic scatter decreases, likely due to the smaller scatter in the MW OC data. Including the data from external galaxies produces a shallower slope ($\beta = 0.229$), likely due to fewer low-mass clusters with small radii in the M31 PHAT sample. Including all data produces a fit similar to the fit for LEGUS + MW, likely due to the leverage and large numbers provided by the low-mass clusters in the MW sample. In all cases, $R_4$ is consistent with that measured in the LEGUS sample alone. Note that due to the asymmetric shape of the cluster radius distributions, $R_4$ may be different from the peak value of the distribution quoted in Section~\ref{sec:radius_dist}.

\subsection{Are clusters gravitationally bound?}

\begin{figure}
    \includegraphics[width=\columnwidth]{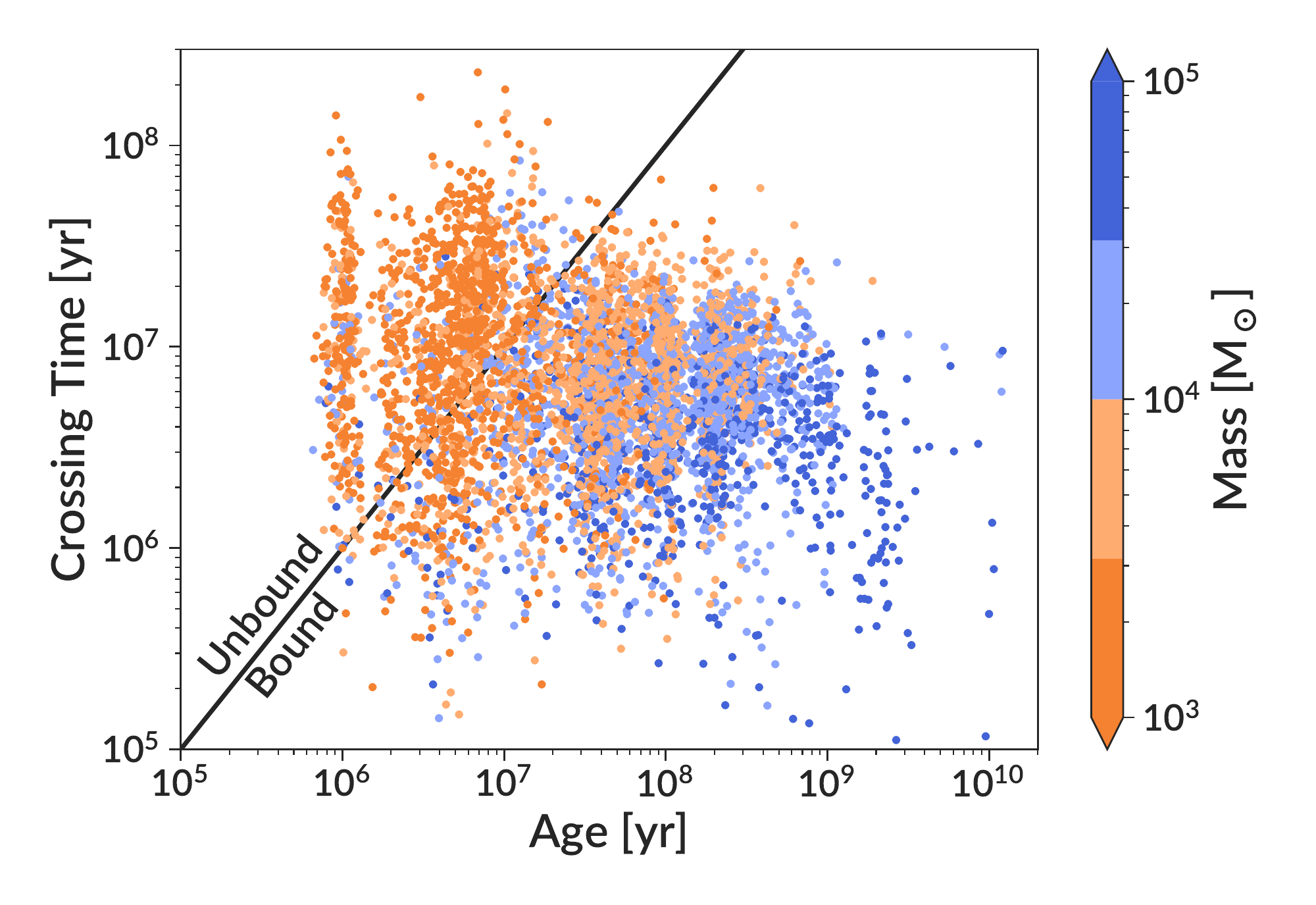}
    \vspace{-5mm}
    \caption{A comparison of the crossing time to the age of clusters in LEGUS. Clusters are color coded according to their mass, and a small random offset was added to the discrete ages for visual purposes. The black line shows where these times are equal. Clusters where the age is longer than the crossing time are likely bound, while those where the age is smaller are likely unbound.}
    \label{fig:crossing_time}
\end{figure}

\begin{figure}
    \includegraphics[width=\columnwidth]{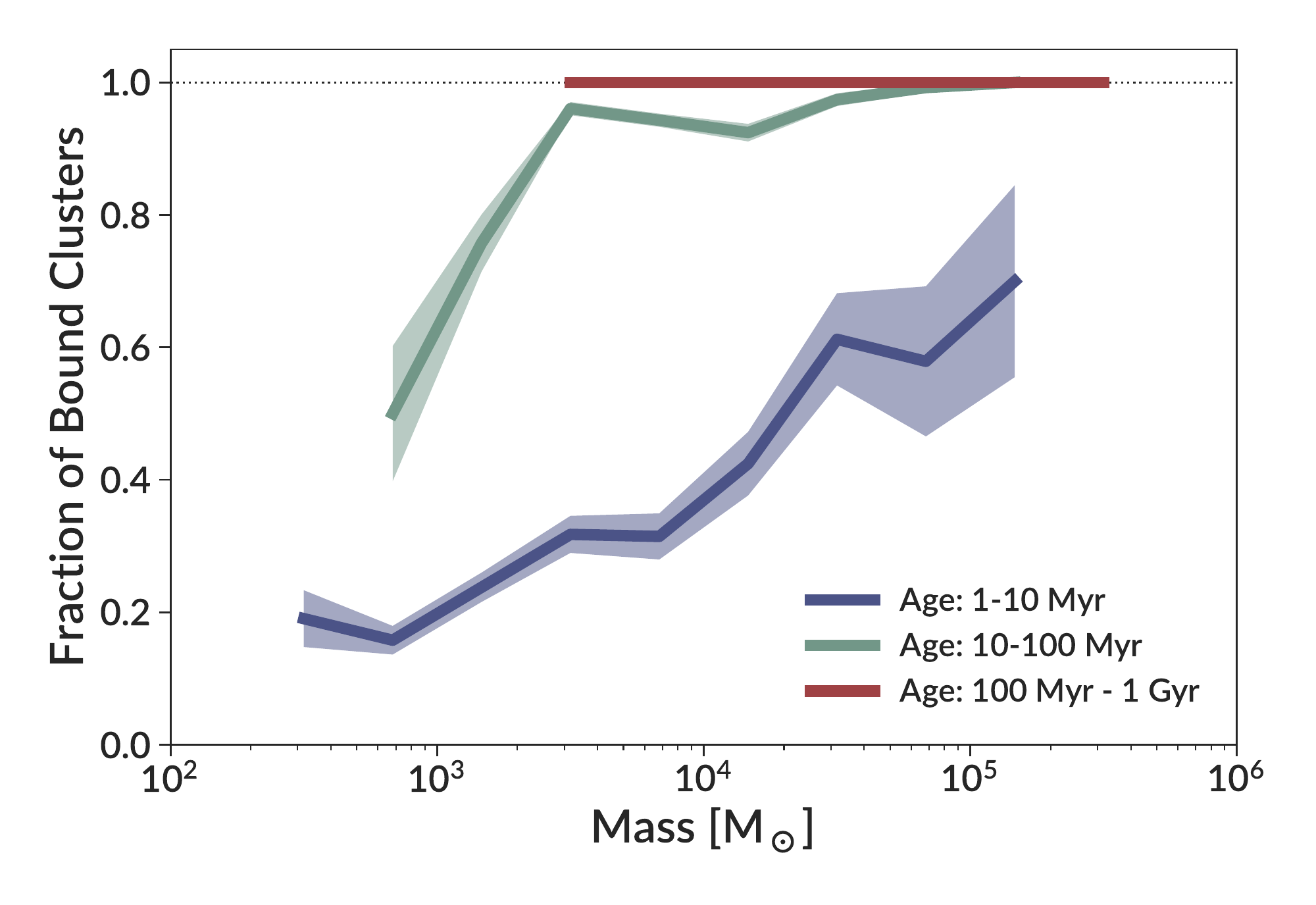}
    \vspace{-5mm}
    \caption{The fraction of clusters that are older than their crossing times (indicating that they are bound objects) as a function of mass. Shaded regions show the 68\% confidence region.}
    \label{fig:bound_fraction}
\end{figure}

\citet{gieles_portegies_zwart_11} suggest a distinction between bound clusters and unbound associations, where bound clusters are older than their instantaneous crossing time:
\begin{equation}
    t_\mathrm{cross} = 10 \left(\frac{\reff^3}{GM}\right)^{1/2}
\end{equation}
Clusters with $t_\mathrm{age} > t_\mathrm{cross}$ have remained together for their lifetimes, indicating that they are gravitationally bound. Unbound objects expand with time, causing the crossing time to increase proportionally.

Figure~\ref{fig:crossing_time} shows the comparison of these timescales for the clusters in the LEGUS sample. The majority of clusters are bound. The only unbound clusters are young (less than 10~Myr) and tend to be less massive. At a given age, the less massive clusters are more likely to be unbound. We find that 78\% of all clusters, 92\% of clusters with $M > 5000 \Msun$, and 97\% of clusters older than 10~Myr are bound. 

Figure~\ref{fig:bound_fraction} shows how the fraction of clusters that are bound changes with age and mass. 100\% of objects older than 100~Myr are bound, while in the other age bins a clear trend with mass is seen. For the 10--100~Myr bin, nearly all clusters above $3000 \Msun$ are bound, while the youngest clusters show a steadily increasing fraction of bound clusters with mass. This confirms that the LEGUS pipeline selects gravitationally bound objects, especially for clusters with higher masses or older ages.

\section{Discussion}
\label{sec:discussion}
\subsection{Selection effects}
\label{sec:selection}

When selecting clusters, the LEGUS survey used an absolute magnitude cut, selecting clusters with a $V$ band magnitude brighter than $-6$ mag. As massive stars in clusters die, the cluster fades. This means that older clusters must be more massive to be detected, producing a significant selection effect in the LEGUS sample. 

This is particularly visible in Figure~\ref{fig:mass_radius_agesplit}. In the oldest age bin, there are nearly no clusters below a few $10^3 \Msun$, where the bulk of the youngest clusters are. The mass ranges seen in Figure~\ref{fig:mass_radius_agesplit} should not be simply interpreted as evidence for disruption of low-mass clusters, as old low-mass clusters would not be detected even if they existed. This also complicates an examination of cluster evolution, as without old low-mass clusters to compare, it is difficult to test predictions of low-mass cluster evolution. 

In the results above and discussion that follows, we present results using all LEGUS clusters. We want to be clear that this is not a complete sample. Where relevant, we discuss how these selection effects may bias the results presented.

LEGUS also uses a cut in concentration index (with a value that varies for each galaxy). This cut may result in the removal of the smallest clusters. This will depend on galaxy distance, as the smallest clusters will be possible to resolve in nearby galaxies but not distant ones. As mentioned above in Section~\ref{sec:radius_dist}, this is not likely to affect many of our galaxies. \citet{Adamo_etal_17} examined the completeness as a function of radius for NGC~628 at a distance of 8.8~Mpc, finding that LEGUS includes roughly 50\% of clusters with $\reff=1$~pc. As most of our galaxies are closer than NGC~628, they will be less affected. The radius distribution shows a clear peak significantly above the radius where we may be incomplete (Figure~\ref{fig:radius_dist}), showing that the potential removal of small clusters likely will not dramatically change our results. 

We also note that because of the inability of our pipeline to pick up extremely small objects (smaller than 0.3 pixels; see Figure~\ref{fig:artificial_clusters}), the smallest objects may be even more compact than reported. 

The inclusion of small, low-mass clusters would have the effect of steepening the mass-radius relation. Interestingly, a comparison of the slopes in Figures~\ref{fig:mass_radius} and \ref{fig:mass_radius_other_samples} shows that a steeper slope better matches the MW OCs. If the true mass-radius relation is steeper than we measure here, it may make our results more consistent with the measurements in the MW. 

Lastly, we note that the radial coverage of each galaxy varies. About half of the LEGUS galaxies are compact enough that they can be covered with one HST WFC3/UVIS pointing, but larger galaxies are not completely covered. Some galaxies include only central regions (e.g. NGC~1566), while others include the central regions and part of the disk (e.g. NGC~628). See Figure~3 of \citet{calzetti_etal_15_legus} for the full footprints for all LEGUS galaxies. This uneven coverage may bias our results somewhat if cluster populations vary throughout galaxies. This could happen if they are tidally bound, as the tidal radius would change with galactocentric radius. We defer a detailed examination of this for a future paper.

\subsection{Mass-radius relation}

The mass-radius relation shown in Figure~\ref{fig:mass_radius} has a relatively shallow slope and significant intrinsic scatter. Nevertheless, a relation is clearly present, even when splitting by age (Figure~\ref{fig:mass_radius_agesplit}).

In the Milky Way, observations of giant molecular clouds (GMCs) show roughly constant surface densities \citep{larson_81}. From this we can expect a mass-radius relation $R = (M/\pi \Sigma)^{1/2} \propto M^{1/2}$. Measurements of clumps have found a range of slopes from 0.3 to 0.6 \citep{romanduval_etal_10,urquhart_etal18,mok_etal_21}. These relations are steeper than the relation we measure for young clusters, which presumably form from such clouds. However, we note that the hierarchical structure of the ISM makes determining the size of a clump more challenging than measuring the radius of a star cluster, so these radii might not be directly comparable \citep{colombo_etal_15}. We examine cluster density further in Section~\ref{sec:density}. The analytic model of \citet{choksi_kruijssen_19} also predicts a mass-radius relation of the form $\reff \propto (M/\Sigma_g)^{1/2}$, where $\Sigma_g$ is the gas surface density. After accounting for the fact that massive clusters can only form in high-density environments, they find a lower slope which is more consistent with this work.

Many previous studies have found inconclusive evidence of a correlation between the mass and radius of young clusters. \citet{zepf_etal_99} found a shallow mass-luminosity relation, with later studies finding little evidence of a strong mass-radius relation \citep{bastian_etal_05,scheepmaker_etal_07,bastian_etal_12,ryon_etal_15,ryon_etal_17}. Some studies have found a mass-radius relation for the most massive clusters above $10^6\Msun$ \citep{kisslerpatig_etal_06,bastian_etal_13}, but that mass range is not sampled in our results. The large sample size, uniform LEGUS selection, and uniform fitting procedure presented here allow us to clearly detect this relation. Our result is similar to that of \citet{cuevasotahola_etal_21}, who find a power law slope of $0.29$. However, the normalization of their relation is higher than ours (see Figure~\ref{fig:mass_radius_other_samples}).

Interestingly, we also find less evolution in the cluster radius with age than seen in other studies. \citet{bastian_etal_12} fit the cluster radius distribution as a bivariate function of age and mass, finding that age is the stronger driver of cluster radius than mass. \citet{ryon_etal_15} found a significant age-radius relation. They also found a steepening of the mass-radius relation with time, although their age bins are different from the bins used in this paper. \citet{chandar_etal16} found that clusters with age of 100--400~Myr have radii 4 times larger than clusters of similar mass with ages younger than 10~Myr. Our results stand in contrast, as we find no significant evolution after 10~Myr in the mass ranges probed.


\subsection{Density distribution}
\label{sec:density}

\begin{figure*}
    \includegraphics[width=1.0\textwidth]{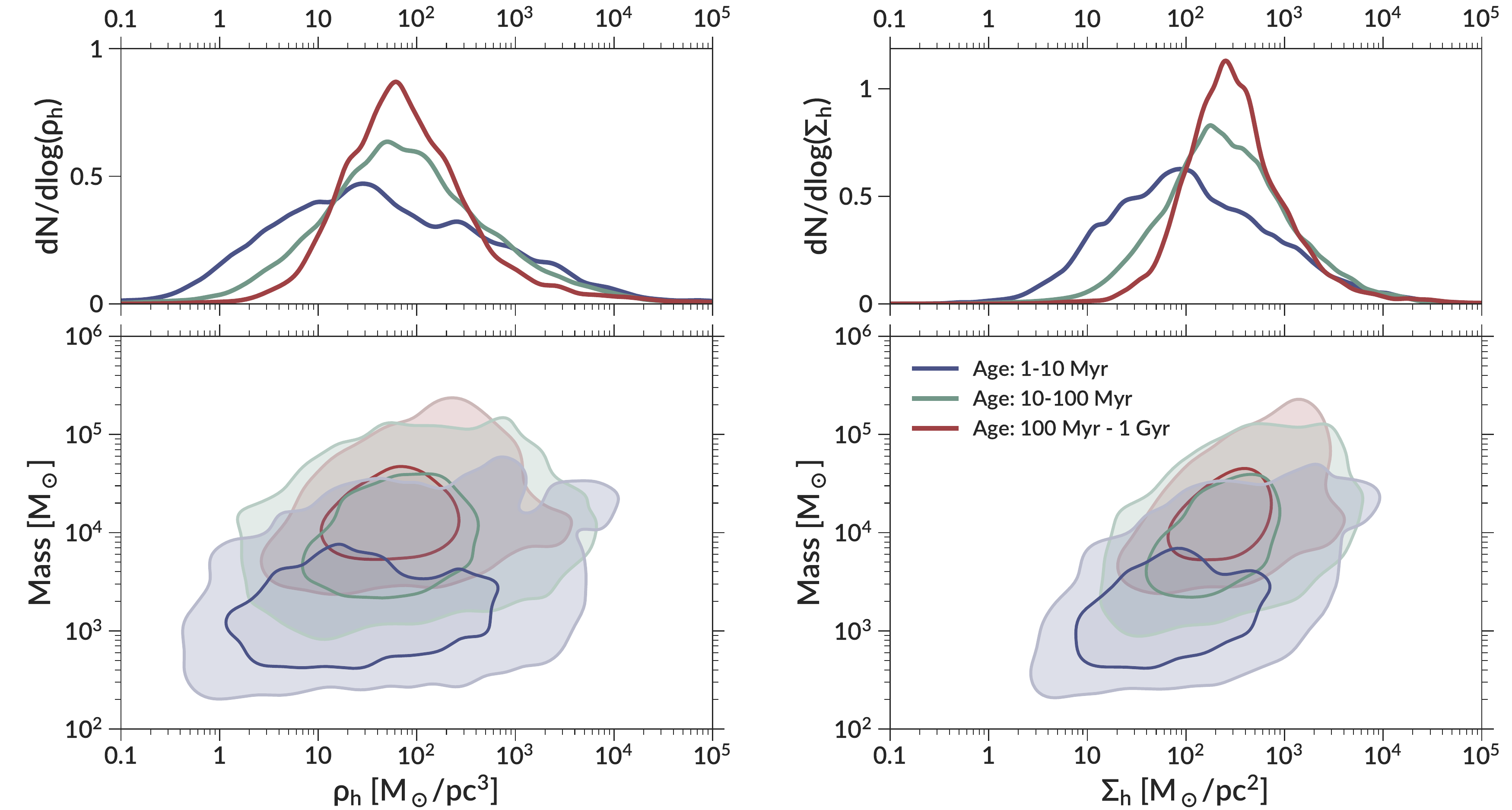}
    \vspace{-5mm}
    \caption{Density $\rho_h$ (left panels) and surface density $\Sigma_h$ (right panels) of clusters within the half-light radius. The top panels shows kernel density estimation of the density distributions, where clusters are smoothed by a Gaussian kernel with a width equal to their measurement error. Each curve is normalized to integrate to the same area. The bottom panels show the dependence of densities on mass. The contours enclose 50 and 90\% of the data, and are smoothed by 0.15 dex. In all panels, we split the sample by age.}
    \label{fig:density}
\end{figure*}

Using our measured radii, we calculate the average density and surface density of the LEGUS clusters:
\begin{equation}
    \rho_h = \frac{3M}{8 \pi \reff^3}, \quad
    \Sigma_h = \frac{M}{2 \pi \reff^2}.
\end{equation}
In this section we will use $\rho_h$ and $\Sigma_h$ when referring to those quantities respectively, and use the generic term ``density'' when referring to both of them. In Figure~\ref{fig:density}, the top panels show the distributions of these densities split by cluster age. Younger clusters have wider ranges and extend to lower densities than old clusters. The bottom panels show the distribution of densities as a function of mass.

We find that the distributions shown in the top panels of Figure~\ref{fig:density} are well described by lognormal functions:
\begin{equation}
    \frac{dN}{d \log \rho_h} = \frac{1}{\sqrt{2 \pi \sigma_\rho^2}} \exp \left[ - \frac{\left( \log \rho_h - \log \mu_\rho \right)^2}{2 \, \sigma_\rho^2} \right]
\end{equation}
and the equivalent for $\Sigma_h$, where $\mu_\rho$ and $\sigma_\rho^2$ are the mean and variance, respectively. We fit these distributions, and show the parameters in Table~\ref{tab:density}. We also include a fit to the entire distribution without splitting by age.

The decrease in the number of low-density clusters with age is likely to be a combination of selection effects and disruption of low-mass clusters. There appears to be a weak mass-$\rho_h$ relation, and a much stronger relation between mass and $\Sigma_h$. As old clusters are massive, they are more likely to have high $\Sigma_h$. However, in the narrow mass range where the age distributions overlap (around $5\times10^4 \Msun$), the youngest age bin extends to lower density than the older age bins. We also examined the density distributions for clusters in the mass range shared by all age bins, again finding a larger spread at younger ages. This may indicate that disruption is responsible for removing these low-density clusters. At the same time, we cannot rule out that higher mass clusters simply form at higher density.

Observations of GMCs in nearby quiescent galaxies are consistent with a roughly constant surface density $\SigmaGMC \sim 100\Msunpc^{-2}$, while in starbursting and high-redshift galaxies the normalization is higher $\SigmaGMC \sim 2000\Msunpc^{-2}$ \citep[e.g.,][]{dessauges_etal19}. Clumps within resolved Galactic clouds are also consistent with a nearly constant surface density $\SigmaGMC \gtrsim 1000\Msunpc^{-2}$ \citep{urquhart_etal18}, although fixed column density may be a selection effect.

The density of the LEGUS clusters is similar to that of GMCs in nearby quiescent galaxies. The peak of the $\Sigma_h$ distribution for young clusters is $\approx100\Msunpc^{-2}$. However, there is a wide range in cluster densities, in contrast to the narrower range of $\SigmaGMC$. This may be due to the fact that there is not a direct connection between the density of GMCs and clusters. Clusters form out of the densest clumps within GMCs, which in the Milky Way typically have $\SigmaGMC$ between 100 and a few $10^4\Msunpc^{-2}$ \citep{urquhart_etal18}. After stars form out of these clumps, stellar feedback disperses the gas. This causes the cluster to increase in size and decrease in density. We note that we are measuring the radii of clusters at this phase of their evolution, after gas expulsion.

We also note that the LEGUS sample is from many galaxies with a range of star formation rates. This may produce a range of GMC properties that are partially responsible for explaining the scatter in cluster density \citep{sun_etal_18}. In future work we will examine the dependence of cluster properties on their environment.

Taking full density distributions from the bottom panels of Figure~\ref{fig:density}, we fit power-law relations and obtain $\rho_h \propto M^{0.52\pm0.02}$ and $\Sigma_h \propto M^{0.67\pm0.012}$. The fitted intrinsic scatter in $\rho_h$ is $1.12\pm0.014$~dex, while for $\Sigma_h$ it is $0.74\pm0.009$~dex. Errors are determined by bootstrapping. As a consistency check, we can compare these fit slopes with those expected based on mass-radius relation fit. For the full LEGUS sample, $\reff \propto M^{0.242}$. By rewriting that relation in terms of densities, the expected relations are $\rho_h \propto M^{0.274}$ and $\Sigma_h \propto M^{0.516}$. These are less steep than the direct fits, especially for $\rho_h$.  Any discrepancy in the slopes may be due to the very large intrinsic scatter in densities. Their dynamic range is larger than the dynamic range in mass. This scatter, along with the lack of a clear relation (particularly for the mass-$\rho_h$ relation), makes it difficult to fit a reliable slope.

\begin{table}
    \centering
    \caption{Lognormal fits to the density distributions shown in Figure~\ref{fig:density}. The log mean $\mu$ and standard deviation $\sigma$ are given for density $\rho_h$ and surface density $\Sigma_h$. Note that the cluster mass ranges given in Table~\ref{tab:fits} apply to these fits too.}
	\begin{tabular}{lcccc}
		\toprule
		Age & $\log \mu_{\rho}$ & $\sigma_{\rho}$ & $\log \mu_{\Sigma}$ & $\sigma_{\Sigma}$ \\ 
		& ($\Msunpc^{-3}$) & (dex) & ($\Msunpc^{-2}$) & (dex) \\ 
		\midrule
		All & 1.80 & 0.78 & 2.36 & 0.61 \\ 
		1--10 Myr & 1.56 & 1.13 & 1.98 & 0.82 \\ 
		10--100 Myr & 1.83 & 0.78 & 2.37 & 0.60 \\ 
		100 Myr -- 1 Gyr & 1.82 & 0.57 & 2.44 & 0.43 \\ 
		\bottomrule
	\end{tabular}
    \label{tab:density}
\end{table}

\subsection{Cluster evolution}
\label{sec:evolution}

Previous literature indicated that clusters may expand with time \citep[e.g.][]{bastian_etal_12,ryon_etal_17}. As stars within clusters lose mass, isolated clusters will slowly expand to maintain virial equilibrium, while at later times two-body relaxation can also increase the cluster radius \citep{gieles_etal_10}.

In Figure~\ref{fig:mass_radius_agesplit} we see a statistically significant evolution in radius with age from the 1--10~Myr bin to the 10--100~Myr bin, with high-mass clusters slightly expanding. However, the magnitude of this increase is quite small. The fit parameters indicate typical clusters at $10^4 \Msun$ expand from 2.36 to 2.51~pc, while clusters at $10^5 \Msun$ expand from 3.58 to 4.76~pc. Notably, we only see this evolution between our first two age bins. We see no significant differences between the 10--100~Myr and 100~Myr--1~Gyr bins. 

To start interpreting these results, we turn to an examination of the Jacobi radius, which sets the radius at which stars belong to a cluster when it is in a tidal field. Clusters that fill a larger fraction of the Jacobi radius are more vulnerable to mass loss. For clusters with mass $M$ in circular orbits with angular frequency $\omega$ in a galaxy with a flat rotation curve, the Jacobi radius is defined as
\begin{equation}
    r_J = \left(\frac{GM}{2 \omega^2}\right)^{1/3}
\end{equation}
We do not directly calculate $r_J$, as obtaining $\omega$ for each cluster is beyond the scope of this paper, and the assumption of a flat rotation curve may not be true for every galaxy.  However, we can qualitatively examine how the ratio of the effective radius to the Jacobi radius scales with cluster mass:
\begin{equation}
    \frac{\reff}{r_J} \propto \frac{M^\beta}{M^{1/3}}
    \label{eq:r_scaling_relation}
\end{equation}
where $\beta$ is the slope of the mass-radius relation (Equation~\ref{eq:mass_radius_relation} and Table~\ref{tab:fits}). Note that this assumes no relation between $M$ and $\omega$. For the full sample $\beta=0.24$, giving $\reff / r_J \propto M^{-0.09}$. In the youngest age bin (1--10~Myr) $\beta=0.18$, and $\reff / r_J \propto M^{-0.15}$. In either case, high-mass clusters fill less of their Jacobi radii than low-mass clusters.

\begin{figure}
    \includegraphics[width=\columnwidth]{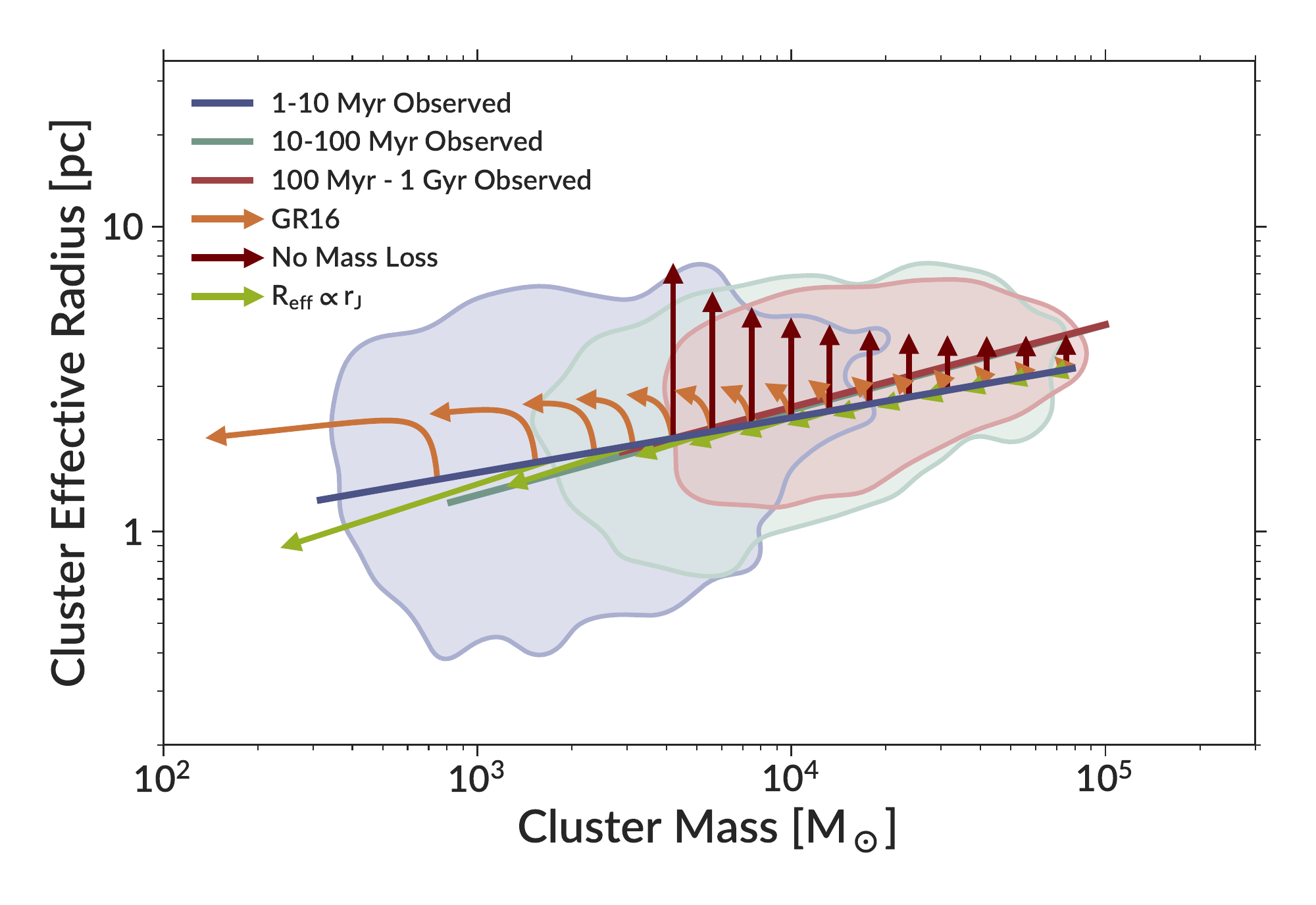}
    \vspace{-5mm}
    \caption{A comparison of 3 models of cluster evolution. Contours enclose 75\% of the clusters in each age bin, and are smoothed by 0.08 dex. The corresponding solid lines show the fit for these age bins from Figure~\ref{fig:mass_radius_agesplit}, but are restricted to the mass range spanned by the 1--99th percentile of masses in this age bin from Table~\ref{tab:fits}. The sets of arrows show how 3 models for cluster evolution as described in Section~\ref{sec:evolution} would change clusters lying on the 1--10~Myr relation after 300~Myr of evolution.}
    \label{fig:toy_model}
\end{figure}

To examine the evolution of clusters with time, we also use the evolution model from \citet{gieles_renaud_16}, known hereafter as GR16. This model includes two processes that influence cluster evolution: tidal shocks and two-body relaxation. Tidal shocks increase the energy of the cluster and result in mass loss, while two-body relaxation is assumed to only increase the energy of the cluster without causing mass loss. To show how this model would change the clusters in the sample, we take clusters along the 1--10~Myr best fit relation and evolve them through this model for 300~Myr. Figure~\ref{fig:toy_model} shows contours of the observed distribution of clusters in the three age bins, their best fit relations, and arrows illustrating how the \citetalias{gieles_renaud_16} model affects clusters in this 300~Myr of evolution. 

Figure~\ref{fig:toy_model} also includes two other toy models of cluster evolution, designed to represent the tidally limited and not tidally limited extremes. When clusters are very small compared to Jacobi radii, any energy injection will increase the radius without causing substantial mass loss. On the other hand, any energy from either two-body relaxation or shocks will cause mass loss in tidally limited clusters, and the cluster's effective radius will decrease along with the tidal radius. 

To model these two cases we modify the mass loss prescriptions of \citetalias{gieles_renaud_16}. They introduce a parameter $f$ that relates mass loss to energy injection (their Equation~2):
\begin{equation}
    \frac{dM}{M} = f \, \frac{dE}{E}
\end{equation}
For shocks, they set $f_\mathrm{sh} = 3$, while setting $f_\mathrm{rlx} = 0$ turns off mass loss from two-body relaxation.

We change these $f$ values to produce our two cases. In the case where clusters are not tidally limited and mass loss does not happen, we set both $f_\mathrm{sh} = f_\mathrm{rlx} = 0$. We then rederive the model, and its results are shown by the ``No Mass Loss'' arrows in Figure~\ref{fig:toy_model}. In the tidally limited case, we keep the original $f_\mathrm{sh} = 3$ but also allow for mass loss from two-body relaxation with $f_\mathrm{rlx} = 0.2$ \citep{gieles_etal06}. As the cluster loses mass, we require the radius to be proportional to the tidal radius:
\begin{equation}
    \frac{\reff}{R_{\mathrm{eff,0}}} =  \frac{r_J}{r_{J,0}} = \left(\frac{M}{M_0}\right)^{1/3}
\end{equation}

The scaling relation presented above in Equation~\ref{eq:r_scaling_relation} indicates that low-mass clusters are more likely to be tidally limited than high-mass clusters, and Figure~\ref{fig:toy_model} qualitatively supports this conclusion. If high-mass clusters are not tidally limited, they will lose little mass and expand, matching the observations. While the mass range of old clusters prohibits a detailed examination of low-mass clusters, the mass-radius relation would steepen if the effective radius of low-mass clusters evolves proportional to the tidal radius.

However, it is clear that none of these models do a good job of quantitatively matching the full evolution. The \citetalias{gieles_renaud_16} model pushes clusters towards a mean relation of $\reff \propto M^{1/9}$, making the relation shallower rather than steeper as required by the observations. The toy model with no mass loss can increase the radius of massive clusters, but its effects are weakest for the highest mass clusters where the observations show the largest difference with age. The toy model that assumes the radius changes with the tidal radius may work for low-mass clusters, but for high-mass clusters it has nearly no effect. 

Importantly, the time dependence of these models is in strong conflict with the observations. We see a change from the 1--10~Myr age bin to the 10--100~Myr bin, with no significant change afterwards. However, the models change clusters steadily with time, leading to little change in the first $\sim$30~Myr but large changes after 300~Myr. 

In addition, models would need to match the change in scatter with time. As clusters evolve towards the mean relation of the \citetalias{gieles_renaud_16} model, the scatter decreases dramatically. We tested this by taking the full 1--10~Myr sample and putting each cluster through the \citetalias{gieles_renaud_16} model for 300~Myr. At this late time, the distribution of cluster radii has a much smaller scatter than seen in the observations of clusters at late times. While the observed scatter does decrease with time, it decreases less than this model predicts. 

We note that stellar mass loss is not included in the \citetalias{gieles_renaud_16} model and the modified versions presented here. In 1 Gyr, clusters can lose roughly 30\% of their mass through stellar evolution alone, and this can cause them to expand \citep{gieles_etal_10}. In addition, one should be careful comparing the cluster mass from the models to the observed mass. The models treat mass loss from stars leaving the cluster as instantaneous, while in reality stars can remain in clusters for a long time before escaping through the Lagrangian points \citep{fukushige_heggie_00,baumgardt_01,claydon_etal_19}. Once stars escape, their low velocity dispersion means that unbound stars can remain near the cluster \citep{kupper_etal_08,kupper_etal_12,webb_etal_13}. These unbound but nearby stars may still be included in the SED fit and radius fits. The observed mass of the cluster is therefore not necessarily the bound mass of the cluster, which is what the models present.

From all of this, it is clear that more work needs to be done to understand how cluster evolution models can be used to interpret this data set. The scaling relation presented above in Equation~\ref{eq:r_scaling_relation} indicates that high-mass clusters may be less likely to be tidally bound, but beyond that we refrain from drawing definitive conclusions about cluster evolution.

\section{Conclusions}
\label{sec:conclusions}

We implemented a custom pipeline to measure the projected half-light radius $\reff$ of star clusters. This pipeline has several features designed to make it robust to contamination and accurately estimate the local background, producing reliable values of $\reff$ (Section~\ref{sec:methods}). We applied this pipeline to the star clusters in 31 LEGUS galaxies, producing a uniformly-measured sample of 7242 star cluster radii. Of these, we identify 6097 as having reliable radii. This is currently the largest such catalog of star cluster radii available. 

We summarize the key results below:

$\bullet$ Most, but not all, galaxies share a common cluster radius distribution, with a peak at around 3~pc (Figure~\ref{fig:radius_dist}). The shape of this distribution is asymmetric, with a tail to small radii, and is well described by a Weibull distribution (Equation~\ref{eq:weibull}).

$\bullet$ We find a clear but shallow mass-radius relation (Figures~\ref{fig:mass_radius},~\ref{fig:mass_radius_other_samples}). This relation takes the form $\reff\propto M^{0.24}$, with an intrinsic scatter of 0.25~dex (Table~\ref{tab:fits}). 

$\bullet$ This mass-radius relation is present in clusters of all ages probed by LEGUS sample (Figure~\ref{fig:mass_radius_agesplit}). The slope of this relation is shallower at 1--10~Myr than at later times, but the slope does not evolve between the 10--100~Myr and 100~Myr--1~Gyr bins. The intrinsic scatter decreases with time. Selection effects cause the subsamples of different age to span different mass ranges, complicating interpretation (Section~\ref{sec:selection}). 

$\bullet$ The majority of clusters identified in LEGUS are gravitationally bound (Figure~\ref{fig:crossing_time}). The majority of unbound clusters are younger than 10~Myr and tend to be less massive (Figure~\ref{fig:bound_fraction}). 

$\bullet$ The distributions of both average density and surface density of LEGUS clusters are well fit by lognormal distributions (Figure~\ref{fig:density}). The width of these distributions is large but decreases with cluster age. The peaks of the distributions for the youngest clusters are $\rho_h\approx30\Msunpc^{-3}$ and $\Sigma_h\approx100\Msunpc^{-2}$ (Table~\ref{tab:density}).

$\bullet$ While we do not directly calculate the Jacobi radii for the LEGUS clusters, the shallow mass-radius relation implies that high-mass clusters fill less of their Jacobi radius than low-mass clusters (Equation~\ref{eq:r_scaling_relation}). 

$\bullet$ We create simple toy models of cluster evolution based on the model in \citet{gieles_renaud_16} to interpret the trends we see with cluster age (Figure~\ref{fig:toy_model}). None of the models can successfully reproduce all aspects of the observed distributions. 

\section*{Acknowledgements}

This research was supported in part by the US National Science Foundation through grants 1909063 and PHY-1748958. We are grateful to the Kavli Institute for Theoretical Physics in Santa Barbara for sponsoring an online program \textit{Globular Clusters at the Nexus of Star and Galaxy Formation} in May--June 2020, which inspired this research.

We thank Arya Farahi for fruitful discussions that shaped our approach to incorporating selection effects in the hierarchical Bayesian model as described in Appendix~\ref{appendix:fitting}. We also thank the referee for suggestions that improved  the paper.

\section*{Data availability}

Cluster catalogs are available on Github at \url{https://gillenbrown.github.io/LEGUS-sizes/}. These catalogs include all quantities needed to reproduce the plots in this paper, including the fit parameters for each cluster. This repository also holds all the code used to fit the radii and generate the figures in this paper. The catalogs include properties such as mass and age that are originally from the LEGUS catalogs, which are found at \url{https://archive.stsci.edu/prepds/legus/dataproducts-public.html}.


\bibliographystyle{mnras}
\bibliography{legus}

\begin{thebibliography}{}
\makeatletter
\relax
\def\mn@urlcharsother{\let\do\@makeother \do\$\do\&\do\#\do\^\do\_\do\%\do\~}
\def\mn@doi{\begingroup\mn@urlcharsother \@ifnextchar [ {\mn@doi@}
  {\mn@doi@[]}}
\def\mn@doi@[#1]#2{\def\@tempa{#1}\ifx\@tempa\@empty \href
  {http://dx.doi.org/#2} {doi:#2}\else \href {http://dx.doi.org/#2} {#1}\fi
  \endgroup}
\def\mn@eprint#1#2{\mn@eprint@#1:#2::\@nil}
\def\mn@eprint@arXiv#1{\href {http://arxiv.org/abs/#1} {{\tt arXiv:#1}}}
\def\mn@eprint@dblp#1{\href {http://dblp.uni-trier.de/rec/bibtex/#1.xml}
  {dblp:#1}}
\def\mn@eprint@#1:#2:#3:#4\@nil{\def\@tempa {#1}\def\@tempb {#2}\def\@tempc
  {#3}\ifx \@tempc \@empty \let \@tempc \@tempb \let \@tempb \@tempa \fi \ifx
  \@tempb \@empty \def\@tempb {arXiv}\fi \@ifundefined
  {mn@eprint@\@tempb}{\@tempb:\@tempc}{\expandafter \expandafter \csname
  mn@eprint@\@tempb\endcsname \expandafter{\@tempc}}}

\bibitem[\protect\citeauthoryear{{Adamo}, {{\"O}stlin}, {Zackrisson}, {Hayes},
  {Cumming}  \& {Micheva}}{{Adamo} et~al.}{2010}]{adamo_etal_10}
{Adamo} A.,  {{\"O}stlin} G.,  {Zackrisson} E.,  {Hayes} M.,  {Cumming} R.~J.,
   {Micheva} G.,  2010, \mn@doi [\mnras] {10.1111/j.1365-2966.2010.16983.x},
  \href {https://ui.adsabs.harvard.edu/abs/2010MNRAS.407..870A} {407, 870}

\bibitem[\protect\citeauthoryear{{Adamo} et~al.,}{{Adamo}
  et~al.}{2017}]{Adamo_etal_17}
{Adamo} A.,  et~al., 2017, \mn@doi [\apj] {10.3847/1538-4357/aa7132}, \href
  {https://ui.adsabs.harvard.edu/abs/2017ApJ...841..131A} {841, 131}

\bibitem[\protect\citeauthoryear{{Adamo} et~al.,}{{Adamo}
  et~al.}{2020}]{adamo_etal20}
{Adamo} A.,  et~al., 2020, \mn@doi [\ssr] {10.1007/s11214-020-00690-x}, \href
  {https://ui.adsabs.harvard.edu/abs/2020SSRv..216...69A} {216, 69}

\bibitem[\protect\citeauthoryear{{Anderson} \& {King}}{{Anderson} \&
  {King}}{2000}]{anderson_etal_00}
{Anderson} J.,  {King} I.~R.,  2000, \mn@doi [\pasp] {10.1086/316632}, \href
  {https://ui.adsabs.harvard.edu/abs/2000PASP..112.1360A} {112, 1360}

\bibitem[\protect\citeauthoryear{{Astropy Collaboration} et~al.,}{{Astropy
  Collaboration} et~al.}{2013}]{astropy_i}
{Astropy Collaboration} et~al., 2013, \mn@doi [\aap]
  {10.1051/0004-6361/201322068}, \href
  {https://ui.adsabs.harvard.edu/abs/2013A&A...558A..33A} {558, A33}

\bibitem[\protect\citeauthoryear{{Astropy Collaboration} et~al.,}{{Astropy
  Collaboration} et~al.}{2018}]{astropy_ii}
{Astropy Collaboration} et~al., 2018, \mn@doi [\aj] {10.3847/1538-3881/aabc4f},
  \href {https://ui.adsabs.harvard.edu/abs/2018AJ....156..123A} {156, 123}

\bibitem[\protect\citeauthoryear{{Ballesteros-Paredes}
  et~al.,}{{Ballesteros-Paredes} et~al.}{2020}]{ballesteros_etal20}
{Ballesteros-Paredes} J.,  et~al., 2020, \mn@doi [\ssr]
  {10.1007/s11214-020-00698-3}, \href
  {https://ui.adsabs.harvard.edu/abs/2020SSRv..216...76B} {216, 76}

\bibitem[\protect\citeauthoryear{{Bastian}, {Gieles}, {Lamers}, {Scheepmaker}
  \& {de Grijs}}{{Bastian} et~al.}{2005}]{bastian_etal_05}
{Bastian} N.,  {Gieles} M.,  {Lamers} H.~J.~G.~L.~M.,  {Scheepmaker} R.~A.,
  {de Grijs} R.,  2005, \mn@doi [\aap] {10.1051/0004-6361:20041078}, \href
  {https://ui.adsabs.harvard.edu/abs/2005A&A...431..905B} {431, 905}

\bibitem[\protect\citeauthoryear{{Bastian} et~al.,}{{Bastian}
  et~al.}{2012}]{bastian_etal_12}
{Bastian} N.,  et~al., 2012, \mn@doi [\mnras]
  {10.1111/j.1365-2966.2011.19909.x}, \href
  {https://ui.adsabs.harvard.edu/abs/2012MNRAS.419.2606B} {419, 2606}

\bibitem[\protect\citeauthoryear{{Bastian}, {Cabrera-Ziri}, {Davies}  \&
  {Larsen}}{{Bastian} et~al.}{2013}]{bastian_etal_13}
{Bastian} N.,  {Cabrera-Ziri} I.,  {Davies} B.,   {Larsen} S.~S.,  2013,
  \mn@doi [\mnras] {10.1093/mnras/stt1779}, \href
  {https://ui.adsabs.harvard.edu/abs/2013MNRAS.436.2852B} {436, 2852}

\bibitem[\protect\citeauthoryear{{Baumgardt}}{{Baumgardt}}{2001}]{baumgardt_01}
{Baumgardt} H.,  2001, \mn@doi [\mnras] {10.1046/j.1365-8711.2001.04272.x},
  \href {https://ui.adsabs.harvard.edu/abs/2001MNRAS.325.1323B} {325, 1323}

\bibitem[\protect\citeauthoryear{{Bell} \& {de Jong}}{{Bell} \& {de
  Jong}}{2001}]{bell_dejong_01}
{Bell} E.~F.,  {de Jong} R.~S.,  2001, \mn@doi [\apj] {10.1086/319728}, \href
  {https://ui.adsabs.harvard.edu/abs/2001ApJ...550..212B} {550, 212}

\bibitem[\protect\citeauthoryear{{Bertin} \& {Arnouts}}{{Bertin} \&
  {Arnouts}}{1996}]{bertin_arnouts_96}
{Bertin} E.,  {Arnouts} S.,  1996, \mn@doi [\aaps] {10.1051/aas:1996164}, \href
  {https://ui.adsabs.harvard.edu/abs/1996A&AS..117..393B} {117, 393}

\bibitem[\protect\citeauthoryear{{Bland-Hawthorn} \&
  {Gerhard}}{{Bland-Hawthorn} \& {Gerhard}}{2016}]{bland-hawthorn_gerhard_16}
{Bland-Hawthorn} J.,  {Gerhard} O.,  2016, \mn@doi [\araa]
  {10.1146/annurev-astro-081915-023441}, \href
  {https://ui.adsabs.harvard.edu/abs/2016ARA&A..54..529B} {54, 529}

\bibitem[\protect\citeauthoryear{{Bothwell}, {Kennicutt}  \& {Lee}}{{Bothwell}
  et~al.}{2009}]{bothwell_etal_09}
{Bothwell} M.~S.,  {Kennicutt} R.~C.,   {Lee} J.~C.,  2009, \mn@doi [\mnras]
  {10.1111/j.1365-2966.2009.15471.x}, \href
  {https://ui.adsabs.harvard.edu/abs/2009MNRAS.400..154B} {400, 154}

\bibitem[\protect\citeauthoryear{{Bovy}}{{Bovy}}{2017}]{bovy_17}
{Bovy} J.,  2017, \mn@doi [\mnras] {10.1093/mnrasl/slx027}, \href
  {https://ui.adsabs.harvard.edu/abs/2017MNRAS.468L..63B} {468, L63}

\bibitem[\protect\citeauthoryear{Bradley et~al.,}{Bradley
  et~al.}{2019}]{photutils}
Bradley L.,  et~al., 2019, astropy/photutils: v0.7.2,
  \mn@doi{10.5281/zenodo.3568287}, \url
  {https://doi.org/10.5281/zenodo.3568287}

\bibitem[\protect\citeauthoryear{{Calzetti} et~al.,}{{Calzetti}
  et~al.}{2015}]{calzetti_etal_15_legus}
{Calzetti} D.,  et~al., 2015, \mn@doi [\aj] {10.1088/0004-6256/149/2/51}, \href
  {https://ui.adsabs.harvard.edu/abs/2015AJ....149...51C} {149, 51}

\bibitem[\protect\citeauthoryear{{Cantat-Gaudin} et~al.,}{{Cantat-Gaudin}
  et~al.}{2018}]{cantat_gaudin_etal_18}
{Cantat-Gaudin} T.,  et~al., 2018, \mn@doi [\aap]
  {10.1051/0004-6361/201833476}, \href
  {https://ui.adsabs.harvard.edu/abs/2018A&A...618A..93C} {618, A93}

\bibitem[\protect\citeauthoryear{{Cardelli}, {Clayton}  \& {Mathis}}{{Cardelli}
  et~al.}{1989}]{cardelli_etal_89}
{Cardelli} J.~A.,  {Clayton} G.~C.,   {Mathis} J.~S.,  1989, \mn@doi [\apj]
  {10.1086/167900}, \href
  {https://ui.adsabs.harvard.edu/abs/1989ApJ...345..245C} {345, 245}

\bibitem[\protect\citeauthoryear{{Chandar}, {Whitmore}, {Dinino}, {Kennicutt},
  {Chien}, {Schinnerer}  \& {Meidt}}{{Chandar} et~al.}{2016}]{chandar_etal16}
{Chandar} R.,  {Whitmore} B.~C.,  {Dinino} D.,  {Kennicutt} R.~C.,  {Chien}
  L.~H.,  {Schinnerer} E.,   {Meidt} S.,  2016, \mn@doi [\apj]
  {10.3847/0004-637X/824/2/71}, \href
  {https://ui.adsabs.harvard.edu/abs/2016ApJ...824...71C} {824, 71}

\bibitem[\protect\citeauthoryear{{Choksi} \& {Kruijssen}}{{Choksi} \&
  {Kruijssen}}{2019}]{choksi_kruijssen_19}
{Choksi} N.,  {Kruijssen} J.~M.~D.,  2019, arXiv e-prints, \href
  {https://ui.adsabs.harvard.edu/abs/2019arXiv191205560C} {p. arXiv:1912.05560}

\bibitem[\protect\citeauthoryear{{Claydon}, {Gieles}, {Varri}, {Heggie}  \&
  {Zocchi}}{{Claydon} et~al.}{2019}]{claydon_etal_19}
{Claydon} I.,  {Gieles} M.,  {Varri} A.~L.,  {Heggie} D.~C.,   {Zocchi} A.,
  2019, \mn@doi [\mnras] {10.1093/mnras/stz1109}, \href
  {https://ui.adsabs.harvard.edu/abs/2019MNRAS.487..147C} {487, 147}

\bibitem[\protect\citeauthoryear{{Colombo}, {Rosolowsky}, {Ginsburg},
  {Duarte-Cabral}  \& {Hughes}}{{Colombo} et~al.}{2015}]{colombo_etal_15}
{Colombo} D.,  {Rosolowsky} E.,  {Ginsburg} A.,  {Duarte-Cabral} A.,   {Hughes}
  A.,  2015, \mn@doi [\mnras] {10.1093/mnras/stv2063}, \href
  {https://ui.adsabs.harvard.edu/abs/2015MNRAS.454.2067C} {454, 2067}

\bibitem[\protect\citeauthoryear{{Cook} et~al.,}{{Cook}
  et~al.}{2019}]{cook_etal_19}
{Cook} D.~O.,  et~al., 2019, \mn@doi [\mnras] {10.1093/mnras/stz331}, \href
  {https://ui.adsabs.harvard.edu/abs/2019MNRAS.484.4897C} {484, 4897}

\bibitem[\protect\citeauthoryear{{Cuevas-Otahola}, {Mayya}, {Puerari}  \&
  {Rosa-Gonz{\'a}lez}}{{Cuevas-Otahola} et~al.}{2020}]{cuevasotahola_etal_20}
{Cuevas-Otahola} B.,  {Mayya} Y.~D.,  {Puerari} I.,   {Rosa-Gonz{\'a}lez} D.,
  2020, \mn@doi [\mnras] {10.1093/mnras/stz3524}, \href
  {https://ui.adsabs.harvard.edu/abs/2020MNRAS.492..993C} {492, 993}

\bibitem[\protect\citeauthoryear{{Cuevas-Otahola}, {Mayya}, {Puerari}  \&
  {Rosa-Gonz{\'a}lez}}{{Cuevas-Otahola} et~al.}{2021}]{cuevasotahola_etal_21}
{Cuevas-Otahola} B.,  {Mayya} Y.~D.,  {Puerari} I.,   {Rosa-Gonz{\'a}lez} D.,
  2021, \mn@doi [\mnras] {10.1093/mnras/staa3513}, \href
  {https://ui.adsabs.harvard.edu/abs/2021MNRAS.500.4422C} {500, 4422}

\bibitem[\protect\citeauthoryear{{Dessauges-Zavadsky}
  et~al.,}{{Dessauges-Zavadsky} et~al.}{2019}]{dessauges_etal19}
{Dessauges-Zavadsky} M.,  et~al., 2019, \mn@doi [Nature Astronomy]
  {10.1038/s41550-019-0874-0}, \href
  {https://ui.adsabs.harvard.edu/abs/2019NatAs...3.1115D} {3, 1115}

\bibitem[\protect\citeauthoryear{{Elson}, {Fall}  \& {Freeman}}{{Elson}
  et~al.}{1987}]{elson_fall_freeman_87}
{Elson} R. A.~W.,  {Fall} S.~M.,   {Freeman} K.~C.,  1987, \mn@doi [\apj]
  {10.1086/165807}, \href
  {https://ui.adsabs.harvard.edu/abs/1987ApJ...323...54E} {323, 54}

\bibitem[\protect\citeauthoryear{{Foreman-Mackey}, {Hogg}, {Lang}  \&
  {Goodman}}{{Foreman-Mackey} et~al.}{2013}]{foreman_mackey_etal_13}
{Foreman-Mackey} D.,  {Hogg} D.~W.,  {Lang} D.,   {Goodman} J.,  2013, \mn@doi
  [\pasp] {10.1086/670067}, \href
  {https://ui.adsabs.harvard.edu/abs/2013PASP..125..306F} {125, 306}

\bibitem[\protect\citeauthoryear{{Fouesneau} et~al.,}{{Fouesneau}
  et~al.}{2014}]{fouesneau_etal_14}
{Fouesneau} M.,  et~al., 2014, \mn@doi [\apj] {10.1088/0004-637X/786/2/117},
  \href {https://ui.adsabs.harvard.edu/abs/2014ApJ...786..117F} {786, 117}

\bibitem[\protect\citeauthoryear{{Fukushige} \& {Heggie}}{{Fukushige} \&
  {Heggie}}{2000}]{fukushige_heggie_00}
{Fukushige} T.,  {Heggie} D.~C.,  2000, \mn@doi [\mnras]
  {10.1046/j.1365-8711.2000.03811.x}, \href
  {https://ui.adsabs.harvard.edu/abs/2000MNRAS.318..753F} {318, 753}

\bibitem[\protect\citeauthoryear{{Gieles} \& {Portegies Zwart}}{{Gieles} \&
  {Portegies Zwart}}{2011}]{gieles_portegies_zwart_11}
{Gieles} M.,  {Portegies Zwart} S.~F.,  2011, \mn@doi [\mnras]
  {10.1111/j.1745-3933.2010.00967.x}, \href
  {https://ui.adsabs.harvard.edu/abs/2011MNRAS.410L...6G} {410, L6}

\bibitem[\protect\citeauthoryear{{Gieles} \& {Renaud}}{{Gieles} \&
  {Renaud}}{2016}]{gieles_renaud_16}
{Gieles} M.,  {Renaud} F.,  2016, \mn@doi [\mnras] {10.1093/mnrasl/slw163},
  \href {https://ui.adsabs.harvard.edu/abs/2016MNRAS.463L.103G} {463, L103}

\bibitem[\protect\citeauthoryear{{Gieles}, {Portegies Zwart}, {Baumgardt},
  {Athanassoula}, {Lamers}, {Sipior}  \& {Leenaarts}}{{Gieles}
  et~al.}{2006}]{gieles_etal06}
{Gieles} M.,  {Portegies Zwart} S.~F.,  {Baumgardt} H.,  {Athanassoula} E.,
  {Lamers} H.~J.~G.~L.~M.,  {Sipior} M.,   {Leenaarts} J.,  2006, \mn@doi
  [\mnras] {10.1111/j.1365-2966.2006.10711.x}, \href
  {https://ui.adsabs.harvard.edu/abs/2006MNRAS.371..793G} {371, 793}

\bibitem[\protect\citeauthoryear{{Gieles}, {Baumgardt}, {Heggie}  \&
  {Lamers}}{{Gieles} et~al.}{2010}]{gieles_etal_10}
{Gieles} M.,  {Baumgardt} H.,  {Heggie} D.~C.,   {Lamers} H. J.~G.~L.~M.,
  2010, \mn@doi [\mnras] {10.1111/j.1745-3933.2010.00919.x}, \href
  {https://ui.adsabs.harvard.edu/abs/2010MNRAS.408L..16G} {408, L16}

\bibitem[\protect\citeauthoryear{{Goodwin} \& {Bastian}}{{Goodwin} \&
  {Bastian}}{2006}]{goodwin_bastian_06}
{Goodwin} S.~P.,  {Bastian} N.,  2006, \mn@doi [\mnras]
  {10.1111/j.1365-2966.2006.11078.x}, \href
  {https://ui.adsabs.harvard.edu/abs/2006MNRAS.373..752G} {373, 752}

\bibitem[\protect\citeauthoryear{{Grasha}}{{Grasha}}{2018}]{grasha_18}
{Grasha} K.,  2018, PhD thesis, University of Massachusetts

\bibitem[\protect\citeauthoryear{{Grasha} et~al.,}{{Grasha}
  et~al.}{2019}]{grasha_etal_19}
{Grasha} K.,  et~al., 2019, \mn@doi [\mnras] {10.1093/mnras/sty3424}, \href
  {https://ui.adsabs.harvard.edu/abs/2019MNRAS.483.4707G} {483, 4707}

\bibitem[\protect\citeauthoryear{{Hogg}, {Bovy}  \& {Lang}}{{Hogg}
  et~al.}{2010}]{hogg_etal_10}
{Hogg} D.~W.,  {Bovy} J.,   {Lang} D.,  2010, arXiv e-prints, \href
  {https://ui.adsabs.harvard.edu/abs/2010arXiv1008.4686H} {p. arXiv:1008.4686}

\bibitem[\protect\citeauthoryear{{Holtzman} et~al.,}{{Holtzman}
  et~al.}{1992}]{holtzman_etal_92}
{Holtzman} J.~A.,  et~al., 1992, \mn@doi [\aj] {10.1086/116094}, \href
  {https://ui.adsabs.harvard.edu/abs/1992AJ....103..691H} {103, 691}

\bibitem[\protect\citeauthoryear{{Jacobs}, {Rizzi}, {Tully}, {Shaya}, {Makarov}
   \& {Makarova}}{{Jacobs} et~al.}{2009}]{jacobs_etal09}
{Jacobs} B.~A.,  {Rizzi} L.,  {Tully} R.~B.,  {Shaya} E.~J.,  {Makarov} D.~I.,
   {Makarova} L.,  2009, \mn@doi [\aj] {10.1088/0004-6256/138/2/332}, \href
  {https://ui.adsabs.harvard.edu/abs/2009AJ....138..332J} {138, 332}

\bibitem[\protect\citeauthoryear{{Johnson} et~al.,}{{Johnson}
  et~al.}{2012}]{johnson_etal_12}
{Johnson} L.~C.,  et~al., 2012, \mn@doi [\apj] {10.1088/0004-637X/752/2/95},
  \href {https://ui.adsabs.harvard.edu/abs/2012ApJ...752...95J} {752, 95}

\bibitem[\protect\citeauthoryear{{Jord{\'a}n} et~al.,}{{Jord{\'a}n}
  et~al.}{2005}]{jordan_etal_05}
{Jord{\'a}n} A.,  et~al., 2005, \mn@doi [\apj] {10.1086/497092}, \href
  {https://ui.adsabs.harvard.edu/abs/2005ApJ...634.1002J} {634, 1002}

\bibitem[\protect\citeauthoryear{{Kelly}}{{Kelly}}{2007}]{kelly_07}
{Kelly} B.~C.,  2007, \mn@doi [\apj] {10.1086/519947}, \href
  {https://ui.adsabs.harvard.edu/abs/2007ApJ...665.1489K} {665, 1489}

\bibitem[\protect\citeauthoryear{{Kharchenko}, {Piskunov}, {Schilbach},
  {R{\"o}ser}  \& {Scholz}}{{Kharchenko} et~al.}{2013}]{kharchenko_etal_13}
{Kharchenko} N.~V.,  {Piskunov} A.~E.,  {Schilbach} E.,  {R{\"o}ser} S.,
  {Scholz} R.~D.,  2013, \mn@doi [\aap] {10.1051/0004-6361/201322302}, \href
  {https://ui.adsabs.harvard.edu/abs/2013A&A...558A..53K} {558, A53}

\bibitem[\protect\citeauthoryear{{Kissler-Patig}, {Jord{\'a}n}  \&
  {Bastian}}{{Kissler-Patig} et~al.}{2006}]{kisslerpatig_etal_06}
{Kissler-Patig} M.,  {Jord{\'a}n} A.,   {Bastian} N.,  2006, \mn@doi [\aap]
  {10.1051/0004-6361:20054384}, \href
  {https://ui.adsabs.harvard.edu/abs/2006A&A...448.1031K} {448, 1031}

\bibitem[\protect\citeauthoryear{{Kourkchi} \& {Tully}}{{Kourkchi} \&
  {Tully}}{2017}]{kourkchi_tully_17}
{Kourkchi} E.,  {Tully} R.~B.,  2017, \mn@doi [\apj]
  {10.3847/1538-4357/aa76db}, \href
  {https://ui.adsabs.harvard.edu/abs/2017ApJ...843...16K} {843, 16}

\bibitem[\protect\citeauthoryear{{Krumholz} et~al.,}{{Krumholz}
  et~al.}{2015}]{krumholz_etal_15}
{Krumholz} M.~R.,  et~al., 2015, \mn@doi [\apj] {10.1088/0004-637X/812/2/147},
  \href {https://ui.adsabs.harvard.edu/abs/2015ApJ...812..147K} {812, 147}

\bibitem[\protect\citeauthoryear{{Krumholz}, {McKee}  \& {Bland
  -Hawthorn}}{{Krumholz} et~al.}{2019}]{krumholz_etal_19_review}
{Krumholz} M.~R.,  {McKee} C.~F.,   {Bland -Hawthorn} J.,  2019, \mn@doi
  [\araa] {10.1146/annurev-astro-091918-104430}, \href
  {https://ui.adsabs.harvard.edu/abs/2019ARA&A..57..227K} {57, 227}

\bibitem[\protect\citeauthoryear{{K{\"u}pper}, {MacLeod}  \&
  {Heggie}}{{K{\"u}pper} et~al.}{2008}]{kupper_etal_08}
{K{\"u}pper} A. H.~W.,  {MacLeod} A.,   {Heggie} D.~C.,  2008, \mn@doi [\mnras]
  {10.1111/j.1365-2966.2008.13323.x}, \href
  {https://ui.adsabs.harvard.edu/abs/2008MNRAS.387.1248K} {387, 1248}

\bibitem[\protect\citeauthoryear{{K{\"u}pper}, {Lane}  \&
  {Heggie}}{{K{\"u}pper} et~al.}{2012}]{kupper_etal_12}
{K{\"u}pper} A. H.~W.,  {Lane} R.~R.,   {Heggie} D.~C.,  2012, \mn@doi [\mnras]
  {10.1111/j.1365-2966.2011.20242.x}, \href
  {https://ui.adsabs.harvard.edu/abs/2012MNRAS.420.2700K} {420, 2700}

\bibitem[\protect\citeauthoryear{{Larsen}}{{Larsen}}{1999}]{larsen_99}
{Larsen} S.~S.,  1999, \mn@doi [\aaps] {10.1051/aas:1999509}, \href
  {https://ui.adsabs.harvard.edu/abs/1999A&AS..139..393L} {139, 393}

\bibitem[\protect\citeauthoryear{{Larson}}{{Larson}}{1981}]{larson_81}
{Larson} R.~B.,  1981, \mn@doi [\mnras] {10.1093/mnras/194.4.809}, \href
  {https://ui.adsabs.harvard.edu/abs/1981MNRAS.194..809L} {194, 809}

\bibitem[\protect\citeauthoryear{{Lee} et~al.,}{{Lee}
  et~al.}{2009}]{lee_etal_09}
{Lee} J.~C.,  et~al., 2009, \mn@doi [\apj] {10.1088/0004-637X/706/1/599}, \href
  {https://ui.adsabs.harvard.edu/abs/2009ApJ...706..599L} {706, 599}

\bibitem[\protect\citeauthoryear{{Leitherer} et~al.,}{{Leitherer}
  et~al.}{1999}]{leitherer_etal_99_starburst}
{Leitherer} C.,  et~al., 1999, \mn@doi [\apjs] {10.1086/313233}, \href
  {https://ui.adsabs.harvard.edu/abs/1999ApJS..123....3L} {123, 3}

\bibitem[\protect\citeauthoryear{{Li}, {Gnedin}  \& {Gnedin}}{{Li}
  et~al.}{2018}]{li_etal18}
{Li} H.,  {Gnedin} O.~Y.,   {Gnedin} N.~Y.,  2018, \mn@doi [\apj]
  {10.3847/1538-4357/aac9b8}, \href
  {http://adsabs.harvard.edu/abs/2018ApJ...861..107L} {861, 107}

\bibitem[\protect\citeauthoryear{{Mackey} \& {Gilmore}}{{Mackey} \&
  {Gilmore}}{2003a}]{mackey_gilmore_03a}
{Mackey} A.~D.,  {Gilmore} G.~F.,  2003a, \mn@doi [\mnras]
  {10.1046/j.1365-8711.2003.06021.x}, \href
  {https://ui.adsabs.harvard.edu/abs/2003MNRAS.338...85M} {338, 85}

\bibitem[\protect\citeauthoryear{{Mackey} \& {Gilmore}}{{Mackey} \&
  {Gilmore}}{2003b}]{mackey_gilmore_03b}
{Mackey} A.~D.,  {Gilmore} G.~F.,  2003b, \mn@doi [\mnras]
  {10.1046/j.1365-8711.2003.06022.x}, \href
  {https://ui.adsabs.harvard.edu/abs/2003MNRAS.338..120M} {338, 120}

\bibitem[\protect\citeauthoryear{{Ma{\'\i}z Apell{\'a}niz}}{{Ma{\'\i}z
  Apell{\'a}niz}}{2009}]{maiz_apellaniz_09}
{Ma{\'\i}z Apell{\'a}niz} J.,  2009, \mn@doi [\apj]
  {10.1088/0004-637X/699/2/1938}, \href
  {https://ui.adsabs.harvard.edu/abs/2009ApJ...699.1938M} {699, 1938}

\bibitem[\protect\citeauthoryear{{McLaughlin} \& {van der Marel}}{{McLaughlin}
  \& {van der Marel}}{2005}]{McLaughlin_vanderMarel_05}
{McLaughlin} D.~E.,  {van der Marel} R.~P.,  2005, \mn@doi [\apjs]
  {10.1086/497429}, \href
  {https://ui.adsabs.harvard.edu/abs/2005ApJS..161..304M} {161, 304}

\bibitem[\protect\citeauthoryear{{Meurer}, {Heckman}, {Leitherer}, {Kinney},
  {Robert}  \& {Garnett}}{{Meurer} et~al.}{1995}]{meurer_etal_95}
{Meurer} G.~R.,  {Heckman} T.~M.,  {Leitherer} C.,  {Kinney} A.,  {Robert} C.,
   {Garnett} D.~R.,  1995, \mn@doi [\aj] {10.1086/117721}, \href
  {https://ui.adsabs.harvard.edu/abs/1995AJ....110.2665M} {110, 2665}

\bibitem[\protect\citeauthoryear{{Mok}, {Chandar}  \& {Fall}}{{Mok}
  et~al.}{2021}]{mok_etal_21}
{Mok} A.,  {Chandar} R.,   {Fall} S.~M.,  2021, \mn@doi [\apj]
  {10.3847/1538-4357/abe12c}, \href
  {https://ui.adsabs.harvard.edu/abs/2021ApJ...911....8M} {911, 8}

\bibitem[\protect\citeauthoryear{{Olivares E.} et~al.,}{{Olivares E.}
  et~al.}{2010}]{olivares_etal10}
{Olivares E.} F.,  et~al., 2010, \mn@doi [\apj] {10.1088/0004-637X/715/2/833},
  \href {https://ui.adsabs.harvard.edu/abs/2010ApJ...715..833O} {715, 833}

\bibitem[\protect\citeauthoryear{{Peng}, {Ho}, {Impey}  \& {Rix}}{{Peng}
  et~al.}{2002}]{peng_etal02_galfit}
{Peng} C.~Y.,  {Ho} L.~C.,  {Impey} C.~D.,   {Rix} H.-W.,  2002, \mn@doi [\aj]
  {10.1086/340952}, \href
  {https://ui.adsabs.harvard.edu/abs/2002AJ....124..266P} {124, 266}

\bibitem[\protect\citeauthoryear{{Peng}, {Ho}, {Impey}  \& {Rix}}{{Peng}
  et~al.}{2010}]{peng_etal10_galfit}
{Peng} C.~Y.,  {Ho} L.~C.,  {Impey} C.~D.,   {Rix} H.-W.,  2010, \mn@doi [\aj]
  {10.1088/0004-6256/139/6/2097}, \href
  {https://ui.adsabs.harvard.edu/abs/2010AJ....139.2097P} {139, 2097}

\bibitem[\protect\citeauthoryear{{Piskunov}, {Schilbach}, {Kharchenko},
  {R{\"o}ser}  \& {Scholz}}{{Piskunov} et~al.}{2007}]{piskunov_etal_07}
{Piskunov} A.~E.,  {Schilbach} E.,  {Kharchenko} N.~V.,  {R{\"o}ser} S.,
  {Scholz} R.~D.,  2007, \mn@doi [\aap] {10.1051/0004-6361:20077073}, \href
  {https://ui.adsabs.harvard.edu/abs/2007A&A...468..151P} {468, 151}

\bibitem[\protect\citeauthoryear{{Portegies Zwart}, {McMillan}  \&
  {Gieles}}{{Portegies Zwart} et~al.}{2010}]{portegies_zwart_etal_10}
{Portegies Zwart} S.~F.,  {McMillan} S. L.~W.,   {Gieles} M.,  2010, \mn@doi
  [\araa] {10.1146/annurev-astro-081309-130834}, \href
  {https://ui.adsabs.harvard.edu/abs/2010ARA&A..48..431P} {48, 431}

\bibitem[\protect\citeauthoryear{{Roman-Duval}, {Jackson}, {Heyer}, {Rathborne}
   \& {Simon}}{{Roman-Duval} et~al.}{2010}]{romanduval_etal_10}
{Roman-Duval} J.,  {Jackson} J.~M.,  {Heyer} M.,  {Rathborne} J.,   {Simon} R.,
   2010, \mn@doi [\apj] {10.1088/0004-637X/723/1/492}, \href
  {https://ui.adsabs.harvard.edu/abs/2010ApJ...723..492R} {723, 492}

\bibitem[\protect\citeauthoryear{{Ryon} et~al.,}{{Ryon}
  et~al.}{2015}]{ryon_etal_15}
{Ryon} J.~E.,  et~al., 2015, \mn@doi [\mnras] {10.1093/mnras/stv1282}, \href
  {https://ui.adsabs.harvard.edu/abs/2015MNRAS.452..525R} {452, 525}

\bibitem[\protect\citeauthoryear{{Ryon} et~al.,}{{Ryon}
  et~al.}{2017}]{ryon_etal_17}
{Ryon} J.~E.,  et~al., 2017, \mn@doi [\apj] {10.3847/1538-4357/aa719e}, \href
  {https://ui.adsabs.harvard.edu/abs/2017ApJ...841...92R} {841, 92}

\bibitem[\protect\citeauthoryear{{Sabbi} et~al.,}{{Sabbi}
  et~al.}{2018}]{sabbi_etal18}
{Sabbi} E.,  et~al., 2018, \mn@doi [\apjs] {10.3847/1538-4365/aaa8e5}, \href
  {https://ui.adsabs.harvard.edu/abs/2018ApJS..235...23S} {235, 23}

\bibitem[\protect\citeauthoryear{{Scheepmaker}, {Haas}, {Gieles}, {Bastian},
  {Larsen}  \& {Lamers}}{{Scheepmaker} et~al.}{2007}]{scheepmaker_etal_07}
{Scheepmaker} R.~A.,  {Haas} M.~R.,  {Gieles} M.,  {Bastian} N.,  {Larsen}
  S.~S.,   {Lamers} H.~J.~G.~L.~M.,  2007, \mn@doi [\aap]
  {10.1051/0004-6361:20077511}, \href
  {https://ui.adsabs.harvard.edu/abs/2007A&A...469..925S} {469, 925}

\bibitem[\protect\citeauthoryear{{Schlafly} \& {Finkbeiner}}{{Schlafly} \&
  {Finkbeiner}}{2011}]{schlafly_finkbeiner_11}
{Schlafly} E.~F.,  {Finkbeiner} D.~P.,  2011, \mn@doi [\apj]
  {10.1088/0004-637X/737/2/103}, \href
  {https://ui.adsabs.harvard.edu/abs/2011ApJ...737..103S} {737, 103}

\bibitem[\protect\citeauthoryear{{Silva-Villa} \& {Larsen}}{{Silva-Villa} \&
  {Larsen}}{2011}]{silvavilla_larsen_11}
{Silva-Villa} E.,  {Larsen} S.~S.,  2011, \mn@doi [\aap]
  {10.1051/0004-6361/201016206}, \href
  {https://ui.adsabs.harvard.edu/abs/2011A&A...529A..25S} {529, A25}

\bibitem[\protect\citeauthoryear{{Silva-Villa}, {Adamo}, {Bastian}, {Fouesneau}
   \& {Zackrisson}}{{Silva-Villa} et~al.}{2014}]{silvavilla_etal_14}
{Silva-Villa} E.,  {Adamo} A.,  {Bastian} N.,  {Fouesneau} M.,   {Zackrisson}
  E.,  2014, \mn@doi [\mnras] {10.1093/mnrasl/slu028}, \href
  {https://ui.adsabs.harvard.edu/abs/2014MNRAS.440L.116S} {440, L116}

\bibitem[\protect\citeauthoryear{{Spitzer}}{{Spitzer}}{1958}]{spitzer_1958}
{Spitzer} Lyman J.,  1958, \mn@doi [\apj] {10.1086/146435}, \href
  {https://ui.adsabs.harvard.edu/abs/1958ApJ...127...17S} {127, 17}

\bibitem[\protect\citeauthoryear{{Sun} et~al.,}{{Sun}
  et~al.}{2018}]{sun_etal_18}
{Sun} J.,  et~al., 2018, \mn@doi [\apj] {10.3847/1538-4357/aac326}, \href
  {https://ui.adsabs.harvard.edu/abs/2018ApJ...860..172S} {860, 172}

\bibitem[\protect\citeauthoryear{{Tully}, {Courtois}  \& {Sorce}}{{Tully}
  et~al.}{2016}]{tully_etal_16}
{Tully} R.~B.,  {Courtois} H.~M.,   {Sorce} J.~G.,  2016, \mn@doi [\aj]
  {10.3847/0004-6256/152/2/50}, \href
  {https://ui.adsabs.harvard.edu/abs/2016AJ....152...50T} {152, 50}

\bibitem[\protect\citeauthoryear{{Urquhart} et~al.,}{{Urquhart}
  et~al.}{2018}]{urquhart_etal18}
{Urquhart} J.~S.,  et~al., 2018, \mn@doi [\mnras] {10.1093/mnras/stx2258},
  \href {https://ui.adsabs.harvard.edu/abs/2018MNRAS.473.1059U} {473, 1059}

\bibitem[\protect\citeauthoryear{{V{\'a}zquez} \& {Leitherer}}{{V{\'a}zquez} \&
  {Leitherer}}{2005}]{vasquez_leitherer_05_starburst}
{V{\'a}zquez} G.~A.,  {Leitherer} C.,  2005, \mn@doi [\apj] {10.1086/427866},
  \href {https://ui.adsabs.harvard.edu/abs/2005ApJ...621..695V} {621, 695}

\bibitem[\protect\citeauthoryear{{Wagner-Kaiser}, {Sarajedini}, {Dalcanton},
  {Williams}  \& {Dolphin}}{{Wagner-Kaiser}
  et~al.}{2015}]{wagner_kaiser_etal_15}
{Wagner-Kaiser} R.,  {Sarajedini} A.,  {Dalcanton} J.~J.,  {Williams} B.~F.,
  {Dolphin} A.,  2015, \mn@doi [\mnras] {10.1093/mnras/stv880}, \href
  {https://ui.adsabs.harvard.edu/abs/2015MNRAS.451..724W} {451, 724}

\bibitem[\protect\citeauthoryear{{Webb}, {Harris}, {Sills}  \& {Hurley}}{{Webb}
  et~al.}{2013}]{webb_etal_13}
{Webb} J.~J.,  {Harris} W.~E.,  {Sills} A.,   {Hurley} J.~R.,  2013, \mn@doi
  [\apj] {10.1088/0004-637X/764/2/124}, \href
  {https://ui.adsabs.harvard.edu/abs/2013ApJ...764..124W} {764, 124}

\bibitem[\protect\citeauthoryear{{Whitmore} et~al.,}{{Whitmore}
  et~al.}{2010}]{whitmore_etal_10}
{Whitmore} B.~C.,  et~al., 2010, \mn@doi [\aj] {10.1088/0004-6256/140/1/75},
  \href {https://ui.adsabs.harvard.edu/abs/2010AJ....140...75W} {140, 75}

\bibitem[\protect\citeauthoryear{{Zackrisson}, {Rydberg}, {Schaerer},
  {{\"O}stlin}  \& {Tuli}}{{Zackrisson} et~al.}{2011}]{zackrisson_etal_11}
{Zackrisson} E.,  {Rydberg} C.-E.,  {Schaerer} D.,  {{\"O}stlin} G.,   {Tuli}
  M.,  2011, \mn@doi [\apj] {10.1088/0004-637X/740/1/13}, \href
  {https://ui.adsabs.harvard.edu/abs/2011ApJ...740...13Z} {740, 13}

\bibitem[\protect\citeauthoryear{{Zepf}, {Ashman}, {English}, {Freeman}  \&
  {Sharples}}{{Zepf} et~al.}{1999}]{zepf_etal_99}
{Zepf} S.~E.,  {Ashman} K.~M.,  {English} J.,  {Freeman} K.~C.,   {Sharples}
  R.~M.,  1999, \mn@doi [\aj] {10.1086/300961}, \href
  {https://ui.adsabs.harvard.edu/abs/1999AJ....118..752Z} {118, 752}

\makeatother
\end{thebibliography}

\appendix


\section{Methods for Fitting Mass-Radius Relation}
\label{appendix:fitting}

In the main text, we use the orthogonal fitting method described sections 7 and 8 of \citet{hogg_etal_10} (hereafter H10), which we summarize here. In evaluating the Gaussian likelihood of each data point given a linear relation (with parameters of slope $m$ and intercept $b$), we use the displacements and variances projected perpendicular to the line being evaluated. The projected displacement is given by Equation 30 of \citetalias{hogg_etal_10}, and can also be written as:
\begin{equation}
    \Delta_i = \frac{y_i -(m x_i + b)}{\sqrt{1 + m^2}}
\end{equation}
where in our case $x$ is the log of the mass, $y$ is the log of the effective radius, $\beta=m$, and we can calculate $b$ given $\beta$ and our pivot point $R_4$. 

The projected variance is given by Equation 31 of \citetalias{hogg_etal_10}. In our case, the mass and radius errors are independent, so the off-diagonal terms of the covariance matrix are zero, allowing us to simplify that expression:
\begin{equation}
    \sigma^2_i = \frac{m^2 \sigma^2_{x, i} + \sigma^2_{y, i}}{1 + m^2}
\end{equation}
Then we add an intrinsic scatter $\sigma_\mathrm{int}$ orthogonal to the line to the data variance, giving the likelihood for a single data point of 
\begin{equation}
    \mathcal{L}_i =  \frac{1}{\sqrt{2 \pi \left(\sigma^2_i + \sigma_\mathrm{int}^2\right)}} \exp \left( -\frac{\Delta_i^2}{2 (\sigma^2_i + \sigma_\mathrm{int}^2)} \right)
\end{equation}
The total data likelihood is the product of this over all data points. In the main text, we maximize this likelihood to produce final parameter values. 

This method does not incorporate any selection effects. As these do exist in the LEGUS sample, we implemented an additional method to attempt to incorporate those selection effects. 
We use a hierarchical Bayesian model, following \citet{kelly_07}. Each cluster has its observed mass and radius ($\Mobsi$, $\reffi$) with corresponding unobserved true quantities ($m_i$, $r_i$). We use the same relation as the main text (Equation~\ref{eq:mass_radius_relation}), but define it using the unobserved true quantities rather than the observed values:
\begin{equation}
    \hat r_i(m_i) = r_4 \left(\frac{m_i}{10^4 \Msun} \right) ^\beta
    \label{eq:mass_radius_relation_underlying}
\end{equation}
such that the normalizing factor $r_4$ is the underlying effective radius at $10^4 \Msun$. In addition, we include an intrinsic lognormal scatter $\sigma_\mathrm{int}$.

For a given cluster, our data likelihood takes the general form:
\begin{align}
    P(\reffi, \Mobsi, r_i, m_i | r_4, \beta, \sigma_\mathrm{int}) =  &P(\reffi | r_i) \\ \notag &\times P(\Mobsi | m_i) \\ \notag &\times P(r_i | m_i, r_4, \beta, \sigma_\mathrm{int}) \\ \notag &\times P(m_i)
\end{align}
where the first two terms are the likelihoods of the observed values given the unobserved true values (which we treat as independent), the third term is the mass-radius relation, and the final term is the prior on the true mass. We model the radius distribution at a given mass as a lognormal distribution:
\begin{equation}
    P(r_i | m_i, r_4, \beta, \sigma_\mathrm{int}) = \frac{1}{\sigma_\mathrm{int} \sqrt{2 \pi}} \exp \left[ -\frac{1}{2} \left(\frac{\log r_i - \log\hat{r_i}(m_i)}{\sigma_\mathrm{int}}\right)^2  \right] 
\end{equation}
where $\log\hat{r_i}(m_i)$ is from Equation~\ref{eq:mass_radius_relation}. The normalizing factor here is important, as it includes a variable of interest $\sigma_\mathrm{int}$. 

We treat the mass and radius measurement errors as independent lognormal variables, with a width equal to the symmetrized observational uncertainties:
\begin{equation}
    P(\reffi | r_i) \propto \exp{\left[-\frac{1}{2} \left(\frac{\log r_i - \log \reffi}{\sigma_{\reffi, \rm{err}}}\right)^2 \right]}
    \label{eq:r_err}
\end{equation}
\begin{equation}
    P(\Mobsi | m_i) \propto \exp{\left[-\frac{1}{2} \left(\frac{\log m_i - \log \Mobsi}{\sigma_{\Mobsi, \rm{err}}}\right)^2 \right]}
    \label{eq:m_err}
\end{equation}
This allows us to analytically marginalize over the unobserved radius $r_i$.
\begin{align}
    \notag P(\reffi | m_i, r_4, \beta, \sigma_\mathrm{int}) = & \int P(\reffi | r_i) P(r_i | m_i, r_4, \beta, \sigma_\mathrm{int}) dr_i \\
    = & \frac{1}{\sqrt{2 \pi \left(\sigma_\mathrm{int}^2 + \sigma^2_{\reffi, \rm{err}}\right)}} \label{eq:mass_radius_likelihood_term} \\
    & \notag \times \exp \left[ -\frac{1}{2} \frac{\left(\log \reffi - \log\hat{r_i}(m_i)\right)^2}{\sigma_\mathrm{int}^2 + \sigma^2_{\reffi, \rm{err}}}  \right]
\end{align}

We also need to include the selection effects. There are two key selection variables: radius and V band absolute magnitude. 

To prevent contamination from unresolved sources, LEGUS selects clusters by examining the concentration index (CI), the magnitude difference between a 3 pixel and 1 pixel apertures. A hard boundary is drawn: anything above this is a cluster, anything below this is a star \citep{Adamo_etal_17}. 
The V band absolute magnitude cut is simple: selected clusters have absolute $V$ band magnitude brighter than $-6$. Unfortunately, this causes selection effects as a function of both mass and age, as dying massive stars go away, making clusters fade with age. To properly account for this, we need to include both age and V band magnitude into our analysis. In what follows we will use $T, \tau$ for observed and true ages, respectively, and $V, \nu$ for observed and true V band absolute magnitudes.

To renormalize our likelihood function, we need to determine the likelihood of a given cluster of a given true mass and age being selected:
\begin{multline}
    \Phi_i (m_i, \tau_i, r_4, \beta, \sigma_\mathrm{int}) = \\ \int f(\reff) f(V) P(\reff | m_i, r_4, \beta, \sigma_\mathrm{int}) P(V | m_i, \tau_i) d\reff dV
    \label{eq:phi}
\end{multline}
where $f(\reff), f(V)$ are the selection functions. For $V$, this is a simple step function:
\begin{equation}
    f(V) = 
    \begin{cases}
        1 & \text{for } V < -6 \\
        0 & \text{for } V \geq -6
    \end{cases}
\end{equation}
For radius, this selection function is more complicated. \citet{Adamo_etal_17} made a first attempt at this in the NGC~628c field, finding that clusters with effective radii of 2~pc were entirely recovered, while those with 1~pc were recovered roughly 50\% of the time. For this test, we simply represent the selection probability as 
\begin{equation}
    f(\reff) = \max\left(\frac{\reff}{0.05\,\mathrm{arcsec}}, 1 \right)
\end{equation}
where 0.05~arcsec is roughly 2~pc at the distance of NGC~628. This gives a 100\% selection at 2~pc and 50\% selection at 1~pc. While this functional form is likely inaccurate, quantifying the selection effects in more detail is beyond the scope of this paper.

The final two terms in Equation~\ref{eq:phi} are the likelihood of a given radius being selected at that mass (given by Equation~\ref{eq:mass_radius_likelihood_term}) and the likelihood of observing a given V band magnitude given a cluster's mass and age. We obtain these using the Yggdrasil models matching those used in LEGUS \citep{zackrisson_etal_11}. We can then represent the V band term as a normal distribution:
\begin{equation}
    P(V_i | m_i, \tau_i) \propto \exp{\left[ -\frac{1}{2} \left(\frac{V_i - \nu(m_i, \tau_i)}{\sigma_{V, \rm{err}}}\right)^2 \right]}
\end{equation}
where $\nu(m_i, \tau_i)$ is the expected V band magnitude obtained from Yggdrasil, and $\sigma_{V, \rm{err}}$ is the observed uncertainty in V band magnitude. Note that since this term is for an arbitrary cluster with true mass $m$ and true age $\tau$, we use typical errors of 0.04 magnitudes.

We use this selection probability to renormalize the likelihood. $\Phi_i$ is calculated separately for each cluster and used to renormalize that cluster's likelihood. In addition, we need to incorporate $\tau$ into the likelihood as it enters the selection effects: 
\begin{align}
    P(\reffi, & \Mobsi, T_i, m_i, \tau_i | r_4, \beta, \sigma_\mathrm{int}) = \\
    & \Phi_i^{-1}(m_i, \tau_i, r_4, \beta, \sigma_\mathrm{int}) \\
    & \times P(\reffi | m_i, r_4, \beta, \sigma_\mathrm{int}) \\
    & \times P(\Mobsi | m_i) P(T_i | \tau_i) P(m_i) P(\tau_i) 
\end{align}
We treat the age in the same way we treat the mass and radius errors, as a lognormal distribution with a width equal to the symmetrized observational error:
\begin{equation}
    P(T_i | \tau_i) \propto \exp{\left[-\frac{1}{2} \left(\frac{\log\tau_i - \log T_i}{\sigma_{T, \rm{err}}}\right)^2 \right]}
    \label{eq:T_err}
\end{equation}

As this likelihood is for one cluster, the total likelihood is the product of the likelihood for all clusters. We also use Bayes' Theorem to turn this into posterior likelihoods on our parameters:
\begin{multline}
    P(r_4, \beta, \sigma_\mathrm{int} | \reffi, \Mobsi, m_i) \propto P(r_4) P(\beta) P(\sigma_\mathrm{int}) \\ \times \prod_i P(\reffi, \Mobsi, m_i | r_4, \beta, \sigma_\mathrm{int}) 
\end{multline}
We use flat priors on all parameters. The slope $\beta$ is uniform between $-1$ and $1$, the normalization $r_4$ is uniform between 0.01 and 100~pc, and the intrinsic scatter $\sigma$ is uniform between 0 and 1~dex.

We sample this posterior distribution using the \textit{emcee} implementation of MCMC \citep{foreman_mackey_etal_13}. This allows us to easily marginalize over $m_i$ and $\tau_i$ in postprocessing of the MCMC chain.

\begin{figure}
    \includegraphics[width=\columnwidth]{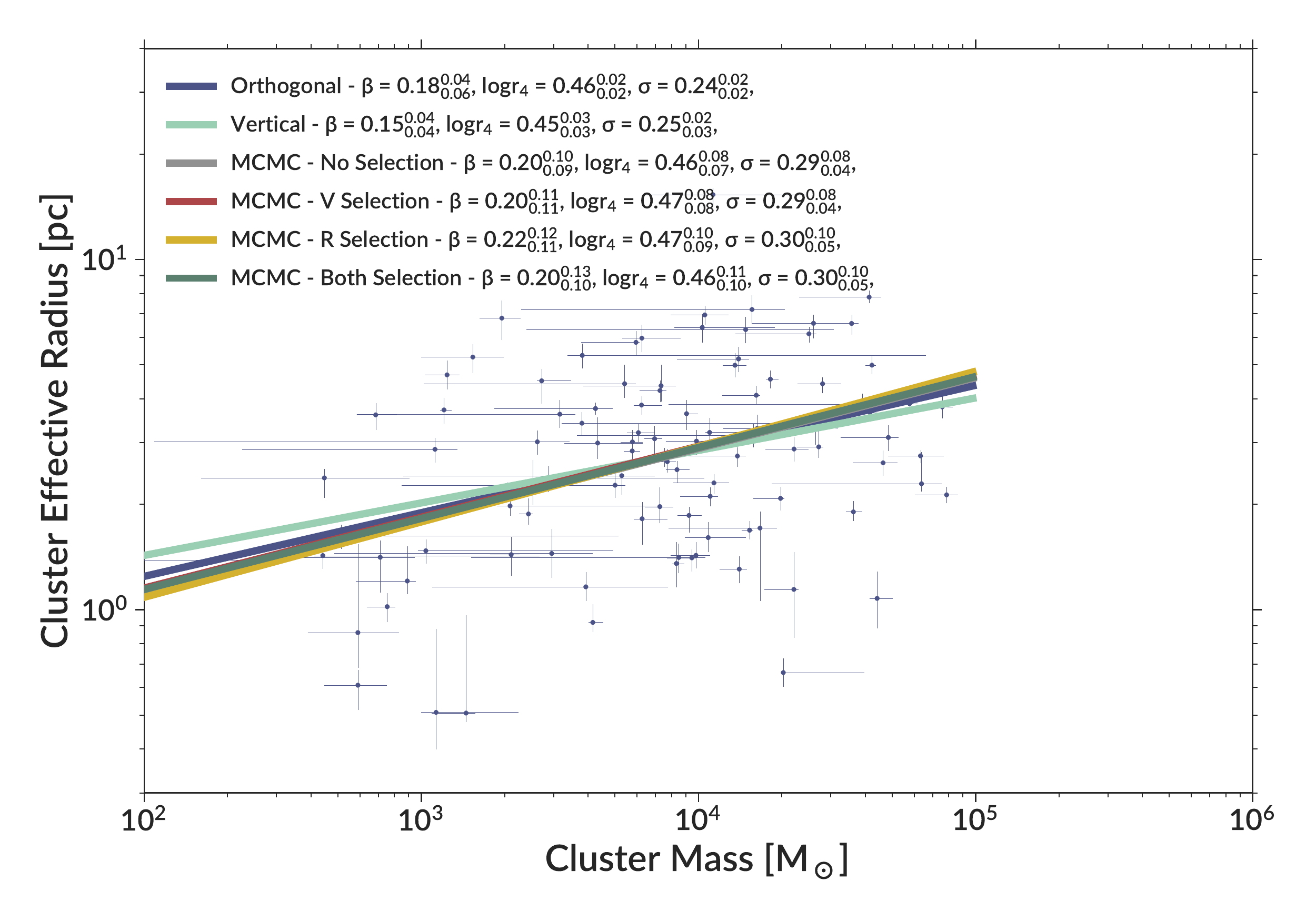}
    \vspace{-5mm}
    \caption{The mass-radius relation fitted by several methods for a random sample of 100 clusters.}
    \label{fig:mass_radius_mcmc}
\end{figure}

As this implementation requires evaluating the integral in Equation~\ref{eq:phi} for each cluster at each step of the MCMC chain, it is computationally expensive, and scales with the number of clusters. We therefore tested this method using a random sample of only 100 clusters. Figure~\ref{fig:mass_radius_mcmc} shows the fit parameters for regular least squares, our fiducial orthogonal least squares method, and several variations of this method with different selection functions enabled.

This hierarchical Bayesian MCMC model produced larger error bars, but otherwise the results are consistent with those of the orthogonal fit. Additionally, removing the selection function terms from the fit does not change the result. Because of this, we decided to not use the hierarchical Bayesian method and instead use the simpler orthogonal fit described at the beginning of this appendix. Another reason to use the orthogonal fit is that we do not know the true selection effects in LEGUS, which would be needed to do a proper analysis of their impact. A full accounting of these is beyond the scope of this paper, making the functional form we assumed for the selection effects overly simplistic. We therefore choose to use the orthogonal fit as our method of choice in this paper.


\bsp	
\label{lastpage}
\end{document}